\input harvmac
\input epsf
\input amssym

\noblackbox
\newcount\figno
\figno=0
\def\fig#1#2#3{
\par\begingroup\parindent=0pt\leftskip=1cm\rightskip=1cm\parindent=0pt
\baselineskip=11pt
\global\advance\figno by 1
\midinsert
\epsfxsize=#3
\centerline{\epsfbox{#2}}
\vskip -21pt
{\bf Fig.\ \the\figno: } #1\par
\endinsert\endgroup\par
}
\def\figlabel#1{\xdef#1{\the\figno}}
\def\encadremath#1{\vbox{\hrule\hbox{\vrule\kern8pt\vbox{\kern8pt
\hbox{$\displaystyle #1$}\kern8pt}
\kern8pt\vrule}\hrule}}

\ifx\pdfoutput\undefined

\else

\fi
%

\def\coeff#1#2{\relax{\textstyle {#1 \over #2}}\displaystyle}

\def\frac#1#2{{#1 \over #2}}

\def\semi{\subset\kern-1em\times\;}

\def\sqr#1#2{{\vcenter{\vbox{\hrule height.#2pt
\hbox{\vrule width.#2pt height#1pt \kern#1pt \vrule width.#2pt}
\hrule height.#2pt}}}}

                  \def\cA{{\cal A}}    
\def\cB{{\cal B}}    
 
               \def\cE{{\cal E}}

 \def\cQ{{\cal Q}}              
                   \def\cS{{\cal S}}       

                   \def\cV{{\cal V}}

\def\ra{{\rightarrow }}

\def\Nb{\overline{N}}

\def\phit{\tilde{\phi}}

%

\def\Nb{\overline{N}}

\def\phit{\tilde{\phi}}

\def\ZZ{\Bbb{Z}}

\def\IR{\Bbb{R}}
\def\kch{k}
\def\lch{\ell}
\def\mch{m}
\def\flux{\Pi}
\def\dist{r}

%
%
%

\lref\vijay{
  V.~Balasubramanian, V.~Jejjala and J.~Simon,
  ``The library of Babel,''
  Int.\ J.\ Mod.\ Phys.\ D {\bf 14}, 2181 (2005)
  [arXiv:hep-th/0505123].
 V.~Balasubramanian, J.~de Boer, V.~Jejjala and J.~Simon,
 ``The library of Babel: On the origin of gravitational thermodynamics,''
  JHEP {\bf 0512}, 006 (2005)
  [arXiv:hep-th/0508023].
}
\lref\MathurHJ{
S.~D.~Mathur, A.~Saxena and Y.~K.~Srivastava,
``Constructing 'hair' for the three charge hole,''
Nucl.\ Phys.\ B {\bf 680}, 415 (2004)
[arXiv:hep-th/0311092].
}
\lref\BenaWT{
  I.~Bena and P.~Kraus,
  ``Three charge supertubes and black hole hair,''
  Phys.\ Rev.\ D {\bf 70}, 046003 (2004)
  [arXiv:hep-th/0402144].
}
\lref\supertube{
D.~Mateos and P.~K.~Townsend,
``Supertubes,''
Phys.\ Rev.\ Lett.\  {\bf 87}, 011602 (2001)
[arXiv:hep-th/0103030].
}
%
\lref\BenaDE{ I.~Bena and N.~P.~Warner, ``One Ring to Rule Them All
... and  in the Darkness Bind Them?,'' arXiv:hep-th/0408106.
}
%
\lref\LuninJY{
O.~Lunin and S.~D.~Mathur,
``AdS/CFT duality and the black hole information paradox,''
Nucl.\ Phys.\ B {\bf 623}, 342 (2002)
[arXiv:hep-th/0109154].
}
%
\lref\MathurZP{
  S.~D.~Mathur,
  ``The fuzzball proposal for black holes: An elementary review,''
  arXiv:hep-th/0502050.
}
%
\lref\GiustoID{
  S.~Giusto, S.~D.~Mathur and A.~Saxena,
  ``Dual geometries for a set of 3-charge microstates,''
  Nucl.\ Phys.\ B {\bf 701}, 357 (2004)
  [arXiv:hep-th/0405017].
}
%
\lref\GiustoKJ{
  S.~Giusto and S.~D.~Mathur,
  ``Geometry of D1-D5-P bound states,''
  arXiv:hep-th/0409067.
}
%
\lref\HorowitzJE{
  G.~T.~Horowitz and H.~S.~Reall,
  ``How hairy can a black ring be?,''
  Class.\ Quant.\ Grav.\  {\bf 22}, 1289 (2005)
  [arXiv:hep-th/0411268].
}
%
\lref\ElvangRT{
  H.~Elvang, R.~Emparan, D.~Mateos and H.~S.~Reall,
  ``A supersymmetric black ring,''
  Phys.\ Rev.\ Lett.\  {\bf 93}, 211302 (2004)
  [arXiv:hep-th/0407065].
}
%
\lref\GauntlettQY{
  J.~P.~Gauntlett and J.~B.~Gutowski,
  ``General concentric black rings,''
  Phys.\ Rev.\ D {\bf 71}, 045002 (2005)
  [arXiv:hep-th/0408122].
}
%
\lref\CyrierHJ{
  M.~Cyrier, M.~Guica, D.~Mateos and A.~Strominger,
  ``Microscopic entropy of the black ring,''
  arXiv:hep-th/0411187.
}
%
\lref\GaiottoXT{
  D.~Gaiotto, A.~Strominger and X.~Yin,
  ``5D Black Rings and 4D Black Holes,''
  arXiv:hep-th/0504126.
}
%
\lref\BenaWV{
  I.~Bena,
  ``Splitting hairs of the three charge black hole,''
  Phys.\ Rev.\ D {\bf 70}, 105018 (2004)
  [arXiv:hep-th/0404073].
}
%
\lref\BenaNI{
  I.~Bena, P.~Kraus and N.~P.~Warner,
``Black Rings in Taub-NUT,''
  arXiv:hep-th/0504142.
}
%
\lref\BenaTK{
  I.~Bena and P.~Kraus,
  ``Microscopic description of black rings in AdS/CFT,''
  JHEP {\bf 0412}, 070 (2004)
  [arXiv:hep-th/0408186].
}
%
\lref\BenaAY{
  I.~Bena and P.~Kraus,
  ``Microstates of the D1-D5-KK system,''
  arXiv:hep-th/0503053.
}
%
\lref\LuninUU{ O.~Lunin, ``Adding momentum to D1-D5 system,'' 
JHEP {\bf 0404}, 054 (2004) [arXiv:hep-th/0404006].
}
%
\lref\GiustoIP{ S.~Giusto, S.~D.~Mathur and A.~Saxena, ``3-charge
geometries and their CFT duals,'' arXiv:hep-th/0406103.
}
%
\lref\ElvangDS{
  H.~Elvang, R.~Emparan, D.~Mateos and H.~S.~Reall,
  ``Supersymmetric black rings and three-charge supertubes,''
  Phys.\ Rev.\ D {\bf 71}, 024033 (2005)
  [arXiv:hep-th/0408120].
}
%
\lref\ElvangSA{
  H.~Elvang, R.~Emparan, D.~Mateos and H.~S.~Reall,
  ``Supersymmetric 4D rotating black holes from 5D black rings,''
  arXiv:hep-th/0504125.
}
%
\lref\BenaTD{
  I.~Bena, C.~W.~Wang and N.~P.~Warner,
  ``Black rings with varying charge density,''
  arXiv:hep-th/0411072.
}
%
\lref\LinNB{
  H.~Lin, O.~Lunin and J.~Maldacena,
``Bubbling AdS space and 1/2 BPS geometries,''
  JHEP {\bf 0410}, 025 (2004)
  [arXiv:hep-th/0409174].
}
\lref\AdSBH{
J.~B.~Gutowski and H.~S.~Reall,
``General supersymmetric AdS(5) black holes,''
JHEP {\bf 0404}, 048 (2004)
 [arXiv:hep-th/0401129].
}
\lref\GutowskiYV{
  J.~B.~Gutowski and H.~S.~Reall,
``General supersymmetric AdS(5) black holes,''
  JHEP {\bf 0404}, 048 (2004)
  [arXiv:hep-th/0401129].
}
%

\lref\LuninFV{
  O.~Lunin and S.~D.~Mathur,
``Metric of the multiply wound rotating string,''
  Nucl.\ Phys.\ B {\bf 610}, 49 (2001)
  [arXiv:hep-th/0105136].
}
%
\lref\BenaJW{
  I.~Bena and N.~P.~Warner,
``A harmonic family of dielectric flow solutions with maximal
supersymmetry,''
  JHEP {\bf 0412}, 021 (2004)
  [arXiv:hep-th/0406145].
}
%
\lref\JejjalaYU{
  V.~Jejjala, O.~Madden, S.~F.~Ross and G.~Titchener,
``Non-supersymmetric smooth geometries and D1-D5-P bound states,''
  arXiv:hep-th/0504181.
}
%

\lref\GibbonsSP{
  G.~W.~Gibbons and P.~J.~Ruback,
``The Hidden Symmetries Of Multicenter Metrics,''
  Commun.\ Math.\ Phys.\  {\bf 115}, 267 (1988).
}
%
\lref\PopeJP{
  C.~N.~Pope and N.~P.~Warner,
``A dielectric flow solution with maximal supersymmetry,''
  JHEP {\bf 0404}, 011 (2004)
  [arXiv:hep-th/0304132].
}
%
\lref\IqbalDS{
  A.~Iqbal, N.~Nekrasov, A.~Okounkov and C.~Vafa,
``Quantum foam and topological strings,''
  arXiv:hep-th/0312022.
}
%
\lref\HawkingZW{
  S.~W.~Hawking,
``Space-Time Foam,''
  Nucl.\ Phys.\ B {\bf 144}, 349 (1978).
}

\lref\PolchinskiUF{
  J.~Polchinski and M.~J.~Strassler,
  ``The string dual of a confining four-dimensional gauge theory,''
  arXiv:hep-th/0003136.
}

%
%

\lref\BenaZB{
  I.~Bena,
  ``The M-theory dual of a 3 dimensional theory with reduced supersymmetry,''
  Phys.\ Rev.\ D {\bf 62}, 126006 (2000)
  [arXiv:hep-th/0004142].
}
\lref\GauntlettNW{
  J.~P.~Gauntlett, J.~B.~Gutowski, C.~M.~Hull, S.~Pakis and H.~S.~Reall,
  ``All supersymmetric solutions of minimal supergravity in five dimensions,''
  Class.\ Quant.\ Grav.\  {\bf 20}, 4587 (2003)
  [arXiv:hep-th/0209114].
}

\lref\KlebanovHB{
  I.~R.~Klebanov and M.~J.~Strassler,
  ``Supergravity and a confining gauge theory: Duality cascades and
  $\chi$SB-resolution of naked singularities,''
  JHEP {\bf 0008}, 052 (2000)
  [arXiv:hep-th/0007191].
}

\lref\VafaWI{
  C.~Vafa,
  ``Superstrings and topological strings at large N,''
  J.\ Math.\ Phys.\  {\bf 42}, 2798 (2001)
  [arXiv:hep-th/0008142].
}
\lref\LuninIZ{
  O.~Lunin, J.~Maldacena and L.~Maoz,
  ``Gravity solutions for the D1-D5 system with angular momentum,''
  arXiv:hep-th/0212210.
}
\lref\GauntlettWH{
  J.~P.~Gauntlett and J.~B.~Gutowski,
  ``Concentric black rings,''
  Phys.\ Rev.\ D {\bf 71}, 025013 (2005)
  [arXiv:hep-th/0408010].
}
\lref\CveticXH{
  M.~Cvetic and F.~Larsen,
  ``Near horizon geometry of rotating black holes in five dimensions,''
  Nucl.\ Phys.\ B {\bf 531}, 239 (1998)
  [arXiv:hep-th/9805097].
}
\lref\GopakumarKI{
  R.~Gopakumar and C.~Vafa,
  ``On the gauge theory/geometry correspondence,''
  Adv.\ Theor.\ Math.\ Phys.\  {\bf 3}, 1415 (1999)
  [arXiv:hep-th/9811131].
}
%
\lref\DenefNB{
  F.~Denef,
``Supergravity flows and D-brane stability,''
  JHEP {\bf 0008}, 050 (2000)
  [arXiv:hep-th/0005049].
}
%
\lref\BatesVX{
  B.~Bates and F.~Denef,
``Exact solutions for supersymmetric stationary black hole composites,''
  arXiv:hep-th/0304094.
}
%
\lref\DenefRU{
  F.~Denef,
``Quantum quivers and Hall/hole halos,''
  JHEP {\bf 0210}, 023 (2002)
  [arXiv:hep-th/0206072].
}
%
\lref\BenaQV{
  I.~Bena and D.~J.~Smith,
  ``Towards the solution to the giant graviton puzzle,''
  Phys.\ Rev.\ D {\bf 71}, 025005 (2005)
  [arXiv:hep-th/0401173].
}
%
\lref\GaiottoGF{
  D.~Gaiotto, A.~Strominger and X.~Yin,
  ``New connections between 4D and 5D black holes,''
  arXiv:hep-th/0503217.
}
%
\lref\KrausGH{
  P.~Kraus and F.~Larsen,
  ``Attractors and black rings,''
  arXiv:hep-th/0503219.
}
%
\lref\MarolfCX{
  D.~Marolf and A.~Virmani,
 ``A black hole instability in five dimensions?,''
  arXiv:hep-th/0505044.
}
\lref\finn{F.~Larsen
``Entropy of Thermally Excited Black Rings,''
 arXiv:hep-th/0505152.
}
\lref\eric{P.~Berglund, E.~G.~Gimon, and T.~S.~Levi, 
``Supergravity Microstates for BPS Black Holes and Black Rings,''
arXiv:hep-th/0505167.
}
%
%
%
%


\Title{
\vbox{
\hbox{\tt hep-th/0505166}\vskip -.15cm
\hbox{\tt UCLA-05-TEP-15}
}}
{\vbox{\vskip -4.5cm
\centerline{\hbox{Bubbling Supertubes and Foaming Black Holes}}}}
\vskip -.3cm
\centerline{Iosif~Bena${}^{(1)}$ and 
Nicholas P.\ Warner${}^{(2)}$}   
\medskip
\centerline{{${}^{(1)}$\it Department of Physics and Astronomy}}
\centerline{{\it University of California}}
\centerline{{\it Los Angeles, CA  90095, USA}}
\medskip
\centerline{{${}^{(2)}$\it Department of Physics and Astronomy}}
\centerline{{\it University of Southern California}}
\centerline{{\it Los Angeles, CA 90089-0484, USA}}
\medskip
\bigskip 
\bigskip
\bigskip
\noindent

We construct smooth BPS three-charge geometries that resolve the
zero-entropy singularity of the $U(1) \times U(1)$ invariant black
ring. This singularity is resolved by a geometric transition that
results in geometries without any branes sources or singularities but
with non-trivial topology.  These geometries are both ground states of
the black ring, and non-trivial microstates of the D1-D5-P system.  We
also find the form of the geometries that result from the geometric
transition of $N$ zero-entropy black rings, and argue that, in general, 
such geometries give a very large number of smooth bound-state
three-charge solutions, parameterized by $6N$ functions.  The generic
microstate solution is specified by a four-dimensional hyper-K\"ahler
geometry of a certain signature, and contains a ``foam'' of
non-trivial two-spheres.  We conjecture that these geometries will
account for a significant part of the entropy of the D1-D5-P black
hole, and that Mathur's conjecture might reduce to counting certain 
hyper-K\"ahler manifolds.

\vskip .3in
\Date{\sl {May, 2005}}

\vfill\eject

\newsec{Introduction}

Mathur and collaborators have proposed a bold solution to the
black hole information paradox \refs{\LuninJY,\MathurHJ,\MathurZP}.  By fully
analyzing the implications of the AdS/CFT correspondence to the physics
of the D1-D5 system, they have argued that each vacuum of the CFT is
dual to a smooth bulk solution that has neither a horizon nor a loss of
information. These geometries thus account for the rather large
entropy of the D1-D5 system. The success of this endeavor for the
D1-D5 system has led to the speculation that one might  similarly
find solutions that account for the entropy of the D1-D5-P system. If
this were possible, then the AdS/CFT correspondence would compel one
to accept that the D1-D5-P black hole should be thought of as an
``ensemble'' of geometries; this would open a new and fascinating
window into the understanding of black holes in string theory.

Most of the progress in understanding whether the D1-D5-P microstates
are dual to bulk solutions has occurred on two apparently distinct
fronts, which this paper will unify.  The first has involved finding
individual smooth solutions carrying D1-D5-P charges, and analyzing
them in the CFT
\refs{\LuninUU\GiustoID\GiustoIP\GiustoKJ{--}\JejjalaYU}.  This has
shown that indeed some CFT microstates are dual to three-charge bulk
geometries, and has highlighted interesting features of the bound
state geometries.

The second has involved understanding the D-brane physics behind the
existence of these solutions, and analyzing these configurations from
a string theory perspective. In particular, in \BenaWT\ it was argued
that there exists a very large class of brane configurations, with
three charges and three dipole charges, that can have arbitrary shape,
and generalize the two-charge supertubes of \supertube. Since the entropy
of the D1-D5 system  comes from the arbitrary
shapes of the two-charge supertubes, it is natural to expect that the
arbitrary shapes of three-charge supertubes account for a sizable part
of the entropy of the D1-D5-P black hole.

Finding the supergravity solutions of these three-charge supertubes of
arbitrary shape is quite involved but in \BenaDE\ it was shown that
one can solve the equations underlying these solutions
\refs{\BenaDE,\AdSBH,\GauntlettNW} in a linear fashion, and reduce the whole
problem of finding three-charge BPS solutions to electromagnetism in
four dimensions.  A side-effect of the study of three-charge
supertubes was the prediction \refs{\BenaWT,\BenaWV} and subsequent
discovery \refs{\ElvangRT,\BenaDE,\ElvangDS,\GauntlettQY} of BPS black rings, 
which by themselves have opened up new windows into 
black-hole physics \refs{\BenaTK\CyrierHJ\BenaTD\HorowitzJE\BenaAY
\GauntlettWH\KrausGH\GaiottoXT\ElvangSA\BenaNI\MarolfCX{--}\finn}. 

For the purpose of finding three-charge geometries dual to
microstates of the D1-D5-P CFT, one is not so much interested in black
rings with a regular event horizon, but rather in the zero-entropy
limit of these rings, which for simplicity we refer to as three-charge
supertubes. The general three-charge supertube solution is given by
six arbitrary functions: four determine the shape of the object, three
describe the charge density profiles but there is one functional
constraint coming from setting the event horizon area to zero
\refs{\BenaDE,\BenaTD}. The near-tube  geometry is of the form AdS$_3 \times S^2$ 
and, since the size of the $S^2$ is finite, the curvature is low
everywhere. However, since the  AdS$_3$ is periodic around the ring, 
these solution have a null orbifold singularity.  In order to
obtain smooth, physical geometries corresponding to supertubes given
by six arbitrary functions, one must learn how to resolve this
singularity.

To do this, we use the fact that singularities coming from wrapped
branes are resolved in string theory via geometric transitions
\refs{\GopakumarKI\KlebanovHB\VafaWI{--}\LinNB}, which result in a
topology change.  The cycle wrapped by the branes shrinks to zero size
while the dual cycle becomes large.  The branes thus disappear from
the space and the naive solution ``transitions'' to one of a different
topology in which the branes have been replaced by a flux through a
non-trivial dual cycle.  After the transition the number of branes is
encoded in the integral of the field strength over this new
topologically non-trivial cycle.  In the limit when the number of
branes become small, the region where the topology change occurs
becomes small and the solution approaches the naive supertube geometry
with its null orbifold singularity.

Unfortunately there is, as yet, no systematic way to find the
geometries that result from a geometric transition of a supertube of 
arbitrary shape. What we can do however is to use the
fact that we know the topology of the solution after the transition
to determine completely the solutions with  $U(1) \times U(1)$
invariance. Having done this, we can easily extend our analysis to 
solutions that only  have (tri-holomorphic) $U(1)$ symmetry, and this leads
to obvious conjectures as to the appropriate backgrounds when there
is no symmetry.

Consider the $\IR^4$ base that contains the supertube/black ring. This
base can be written as a trivial Gibbons-Hawking space with one center
of unit Gibbons-Hawking (GH) charge. The singularity of the supertube is resolved 
by the nucleation of a pair of GH centers, with equal and opposite charges,
$-Q$ and $Q$, near the location of the supertube. Despite the fact
that the signature of the new base can change from $(+,+,+,+)$ to
$(-,-,-,-)$, the overall geometry is regular.  One should also note
that if $|Q| \ne 1$ then there will be ordinary, $\ZZ_{|Q|}$, spatial orbifold singularities 
at the corresponding GH centers.   Such spatial orbifolds are well-understood
in string theory and in the underlying conformal field theory, and are
therefore harmless.   So when we say the solution is regular, we will 
mean up to such spatial orbifolds at the GH centers.  The new solution has a
non-trivial topology, with two two-cycles, but no branes.  A schematic
of this transition is depicted in Fig. 1.  The three
dipole charges of the naive solution are now given by the integrals on
the newly nucleated $S^2$ of the three two-form field strengths on the
base\foot{These two-form field strengths come from reducing the
M-theory four-form field strength on the three $T^2$'s}. The size of
this two-sphere is determined by the balance between the fluxes
wrapping it, and the attraction of the $-Q$ and $Q$ GH centers.  When
the fluxes are very small, or $Q$ very large, these GH centers become
very close, and the solution approaches the naive supertube solution.

\goodbreak\midinsert
\vskip .2cm
\centerline{ {\epsfxsize 4in\epsfbox{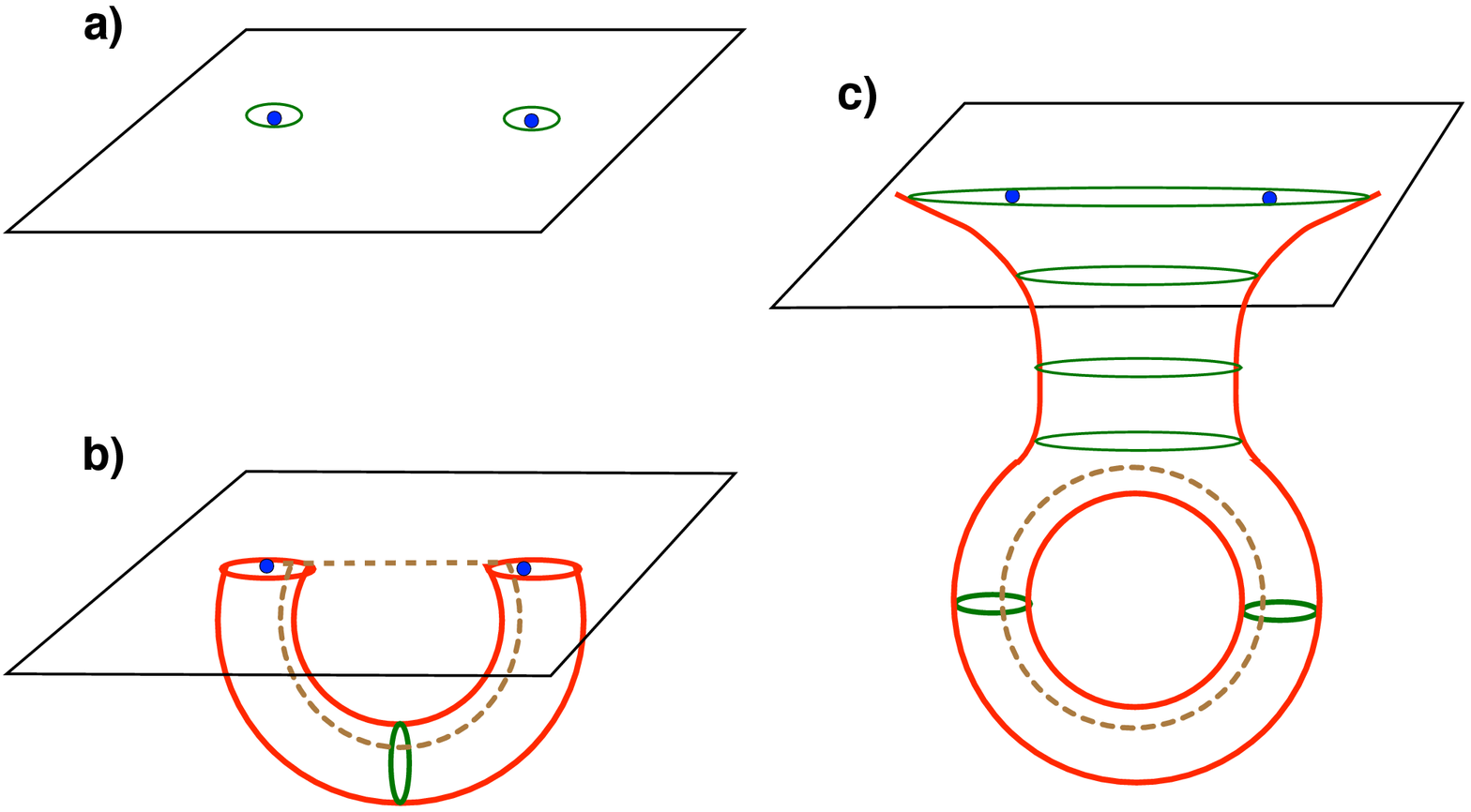}}}
\vskip 0.2cm
\leftskip 2pc
\rightskip 2pc\noindent{\ninepoint\sl \baselineskip=8pt 
{\bf Fig.~1}: The geometric transition: This figure shows a section
through the transition geometry in which an $S^2$ is depicted as an
$S^1$, and the $S^1$ of the supertube as a pair of two blue points. 
The naive geometry is shown in a), and the resolved geometry is shown
in b). After the transition the green  $S^2$ Gaussian 
surface becomes non-contractible and a new two-cycle (depicted 
by the brown dotted $S^1$) appears. In c) we display the transition for large 
dipole charges, when the link with the naive D-brane solution a) becomes much
less obvious.}
\endinsert

The physics of the singularity resolution we propose here is very
similar to, and was inspired by, the one observed in the recent paper
of Lin, Lunin and Maldacena \LinNB, where the bubbling solutions
reduce to naive giant gravitons in the small dipole charge limit, and
have a topological transition in which branes are replaced by fluxes.
As in \LinNB, when the dipole charges become large the bubbling
solution has no obvious brane interpretation.

There are two very non-trivial confirmations that our solutions are
the correct resolving geometries. First, one can put the bubbling
supertube in Taub-NUT, and move it into the four-dimensional
region. The resulting four-dimensional solution is in the same class
as the solutions \BenaAY\ that resolve the singularity of the
zero-entropy four-dimensional black hole.  One can also take the limit
of our solutions in which the $+Q$ GH center is moved onto the center
with a GH charge of $+1$ at the origin.  In this limit, our solutions
reduce to the bound state solutions constructed by Mathur and
collaborators
\refs{\GiustoID,\GiustoIP,\GiustoKJ} by taking novel extremal limits
of the non-extremal rotating three-charge black hole \CveticXH.

Our solutions
suggest quite a few non-trivial features of the three-charge
geometries that are dual to microstates of the three-charge black
hole. We can argue that several concentric rings are resolved by the
nucleation of several pairs of GH centers, one pair for each ring, and
that such a solution is a bound state. This indicates that the most
general bound-state solution with a GH base should have a collection
of GH centers of positive and negative charges at arbitrary positions
inside the $\IR^3$ base of the GH space, with the sum of the charges
equal to one. The solutions have non-vanishing fluxes through the
non-trivial two-cycles of the base, and have no localized brane
charge.

One also expects the geometric transition we present here to resolve
the singularities of the three-charge supertubes of arbitrary
shape. The resolved solutions should have the same topology as the
$U(1)\times U(1)$ solution.  However, their bases will no longer be
multiple-center GH spaces but more general four-dimensional
hyper-K\"ahler manifolds. Even if such base manifolds have changing
signature, we expect the overall solutions to be regular, just as for
the $U(1)\times U(1)$ invariant solutions.  A configuration containing
$N$ supertubes of arbitrary shapes is determined by $6 N$ functions,
and after the transition should give $6N$ functions worth of smooth
geometries. This is a huge number of smooth solutions, which might as
well be large enough to account for a significant part of the entropy
of the three-charge black hole. Is is also possible that a significant
part of this entropy comes from degrees of freedom along the three
$T^2$'s of the solution, which probably cannot be described by 
supergravity.

Our results indicate that proving or disproving  the strong form of Mathur's 
conjecture - that black hole microstates are dual to 
{\it smooth supergravity solutions} \foot{There also exists a weak form of 
Mathur conjecture - that 
black-hole microstates are dual to string theory configurations with unitary 
scattering that are not necessarily smooth in supergravity.} 
- reduces to a well-defined mathematical problem: classifying and counting
asymptotically-flat four-dimensional hyper-K\"ahler manifolds that
have regions of signature $(+,+,+,+)$ and regions of signature
$(-,-,-,-)$. If the conjecture is correct, it indicates that the black
hole is an ``ensemble'' of hyper-K\"ahler geometries involving foams
of a very large number of topologically non-trivial $2$-spheres,
threaded by fluxes.  More generally, our results have potentially
interesting consequences for the structure of the supersymmetric
vacuum states of string theory.  We will discuss this further in the
last section of this paper.

The BPS black rings have two microscopic interpretations: one in terms
of the D1-D5-P CFT \BenaTK\ and another in terms of a four-dimensional
black hole CFT \refs{\BenaTK,\BenaNI,\CyrierHJ}. Hence, our solutions
have two microscopic interpretations.  On one hand, they are dual to
microstates of the black ring CFT, and should be thought of as the
ground states of the BPS black ring, in the same way that the
solutions of \refs{\LuninFV,\LuninIZ} give the ground state of the 
five-dimensional three-charge black hole, and the solutions of \BenaAY\ 
give the ground state of the four-dimensional 
four-charge black hole.  On the other hand, they are dual
to vacua of the D1-D5-P CFT. Our analysis does not establish to which
CFT vacua our solutions are dual, and we leave this very interesting
question to future work. However, based on the microscopic description
of supertubes \refs{\LuninFV,\LuninIZ,\BenaTK} we expect the solutions
that correspond to multiple supertubes to be dual to CFT states with
longer effective strings than the solutions that come from only one
supertube. Hence, the solutions with the largest number of bubbles
should correspond to the CFT states with the longest effective
strings, which are the ones that give the D1-D5-P black hole entropy.

In Section 2 we explain the features of the geometric transition that
resolves the singularity of the zero-entropy black rings. These
features are very similar to those observed in \LinNB\ to give
bubbling AdS geometries. In Section 3 we investigate the general form
of solutions with a Gibbons-Hawking base with an arbitrary
distribution of centers.  In Section 4 we consider solutions in which
there are no point-charge sources and only topologically non-trivial
fluxes.  We also discuss some simple examples.  
The reader who is already familiar with the construction of 
metrics with Gibbons-Hawking base and is primarily interested in the 
solution that resolves the singularity of the zero-entropy black ring
should skip to Section 5.   Indeed, in Section 5 we
explicitly construct the bubbling solutions that come from the
geometric transition of zero-entropy black rings and then compare
their features to those of the naive supertube solutions.  We also
place the bubbling solutions in Taub-NUT and relate them to the ground
states of four-dimensional black holes constructed in \BenaAY.
Section 6 contains some final remarks.

While working on this paper we became aware of another group
that is  working on similar issues \eric.  
Our paper and theirs will appear simultaneously on the archive.
 
\newsec{Singularity Resolution and Geometric Transitions}

As is well known, D-branes warp the geometry by shrinking it along the
longitudinal directions and expanding it along the transverse
directions.  Hence, solutions sourced by branes that wrap a closed
curve generally have a singularity because the tension in the brane
causes this curve to shrink to zero size.  Perhaps the best known
example of such a singularity is Poincar\'e AdS space with a periodic
direction.  For example, if one takes the standard $AdS_5 \times S^5$
solution corresponding to D3-branes, then it is regular but if one
periodically identifies one of the spatial directions of the
D3-branes then those directions collapse to zero size as one goes
down the $AdS$ throat.

An even more interesting example of such a singularity comes from
studying M2 branes polarized into M5 branes by a transverse field
\BenaZB. The M5 branes are wrapped on a topologically trivial $S^3$ and
are stabilized against collapse by the transverse field.  The half-BPS
supergravity solution describing this system \refs{\PopeJP, \BenaJW}
looks like AdS$_4 \times S^7$ far away from the polarization shell,
and like AdS$_7\times S^4$ near the shell. However, because the M5
branes that source the AdS$_7\times S^4$ solution are wrapped on a
three-sphere, the naive near-shell geometry has a singularity.

In \LinNB\ it was realized that this singularity is resolved by a
geometric transition.  The $S^3$ wrapped by the branes shrinks to zero 
size  at the position of  the branes.  The $S^4$ that links this $S^3$ is
a ``Gaussian surface'' for the M5-brane charge and necessarily becomes 
large. Moreover, since the $S^3$ also shrinks to
zero size at the origin of the space, this results in the creation of
another topologically non-trivial $S^4$. The integral of the four-form
flux on the second $S^4$ is equal to $N_{2} \over N_{5}$. Hence,
before the transition the solution had a shell of M2 branes polarized
into M5 branes, and after the transition the solution has a
non-trivial topology, two $S^4$'s threaded by fluxes $N_{5}$ and
$N_{2} \over N_{5}$, and no branes. The M2 brane charge measured at
infinity comes from the non-trivial fluxes through the two $S^4$'s;
these fluxes combine via the supergravity Chern-Simons term to
generate the electric charge.


\goodbreak\midinsert

\def\nicespacea#1{{~~#1~~}}
\goodbreak
{\vbox{\ninepoint{
$$
\vbox{\offinterlineskip\tabskip=0pt
\halign{\strut\vrule#
&\vrule\hfil #\hfil 
&\vrule\hfil #\hfil 
&\vrule\hfil #\hfil 
&\vrule\hfil #\hfil 
&\vrule\hfil #\hfil 
&\vrule\hfil #\hfil 
&\vrule\hfil #\hfil 
&\vrule\hfil #\hfil 
&\vrule\hfil #\hfil 
&\vrule\hfil #\hfil 
&\vrule\hfil #\hfil 
&\vrule\hfil #\hfil 
&\vrule\hfil #\hfil 
&\vrule\hfil #\hfil 
\cr
\noalign{\hrule}
&~~~~~&\nicespacea{1}&\nicespacea{2}&\nicespacea{3}&\nicespacea{4}
&\nicespacea{5}&\nicespacea{6}&\nicespacea{7}&\nicespacea{8}
&\nicespacea{9}&\nicespacea{10}&\nicespacea{11}& \cr
\noalign{\hrule height1pt}
&~M2~&\nicespacea{$|$}&\nicespacea{$|$}&\nicespacea{$|$}&{---}&{---}&{---}&{---}
&\nicespacea{$\star$}&\nicespacea{$\star$}&\nicespacea{$\star$}
&\nicespacea{$\star$}&  \cr
\noalign{\hrule}
&~M2~&\nicespacea{$|$}&{---}&{---}&\nicespacea{$|$}&\nicespacea{$|$}&{---}&{---}
&\nicespacea{$\star$}&\nicespacea{$\star$}&\nicespacea{$\star$}
&\nicespacea{$\star$}&  \cr
\noalign{\hrule}
&~M2~&\nicespacea{$|$}&{---}&{---}&{---}&{---}&\nicespacea{$|$}
&\nicespacea{$|$}
&\nicespacea{$\star$}&\nicespacea{$\star$}&\nicespacea{$\star$}&
\nicespacea{$\star$}& \cr
\noalign{\hrule height1pt}
&~M5~&\nicespacea{$|$}&{---}&{---}&\nicespacea{$|$}&\nicespacea{$|$}
&\nicespacea{$|$}&\nicespacea{$|$}
& &\omit &\omit  $x^\mu (\phi)$  &\omit   &  \cr
\noalign{\hrule}
&~M5~&\nicespacea{$|$}&\nicespacea{$|$}&\nicespacea{$|$}&{---}&{---}
&\nicespacea{$|$}&\nicespacea{$|$}
& &\omit &\omit $x^\mu (\phi)$ &\omit &  \cr
\noalign{\hrule}
&~M5~&\nicespacea{$|$}&\nicespacea{$|$}&\nicespacea{$|$}
&\nicespacea{$|$}&\nicespacea{$|$}&{---}&{---}
& &\omit &\omit $x^\mu (\phi)$  &\omit   &  \cr
\noalign{\hrule}}
\hrule}$$
\vskip-7pt
\leftskip 2pc
\rightskip 2pc\noindent{\ninepoint\sl \baselineskip=8pt
{\bf Table 1}: Layout of the branes that give the supertubes and black rings 
in an M-theory duality frame. Vertical bars
$|$, indicates the directions along which the branes are extended, and 
horizontal lines, ---, indicate the smearing directions. The functions, $x^\mu(\phi)$, 
indicate that the brane wraps a simple closed curve that gives 
the supertube profile. A $\star$ indicates that a brane is smeared along the
supertube profile, and pointlike on the other three directions.}
}
\vskip7pt}}

\endinsert

The three-charge supertubes that one obtains from the zero-entropy
limit of black rings also have a similar singularity. The brane
content of these supertubes is shown in Table 1.  The tubes are
``wrapped'' on a topologically trivial $S^1$ that sits in the spatial
$\IR^4$ base. The integral of the field strength $F_{23ij} dx^i \wedge
dx^j$ on the $S^2$ that surrounds this $S^1$ gives the number of M5
branes that wrap the $4567$ directions and extend along the $S^1$ of
the tube.  Similarly, the integrals of $F_{45ij} dx^i \wedge dx^j$ and
$F_{67ij} dx^i \wedge dx^j$ measure the other two dipole charges of
the solution.  After the geometric transition, the $S^1 \subset \IR^4$
of the tube shrinks to zero size, and the $S^2$ around the supertube
becomes fat.  Moreover, since this $S^1$ also shrinks to zero size at
the origin of $\IR^4$, this will give another topologically
non-trivial $S^2$.  The resulting four-geometry, ${\cal M}^4$, will therefore 
have two non-contractible two-spheres, $S^2_A$ and $S^2_B$, and no brane
sources.  The product of the integrals of the fluxes over these
non-trivial two-spheres, $S^2_A$ and $S^2_B$, will give the M2-brane charges
measured at infinity and induced through the supergravity Chern-Simons term.  
For example, the M2 brane charge along the $23$ directions should be given
by:
\eqn\chern{ Q^{M2}_{23}  ~=~ \half\,  \int_{{\cal M}^4 \times T^2_{45} 
 \times T^2_{67}  }\, F \, \wedge \, F ~=~   {\cal I}^{-1}_{AB}\int_{S^2_A} 
 F_{45ij}\times\int_{S^2_B} F_{67ij} \,,}
where $ {\cal I}_{AB} $ is the intersection matrix of the cycles $A$
and $B$.  After the transition there are no more brane sources and so
the solution should be completely determined by the base space and by
the fluxes. Moreover, in order for the solution to preserve the same
supersymmetries as three sets of M2 branes the base space must be
hyper-K\"ahler \refs{\GauntlettNW,\BenaDE,\AdSBH}.  For the transition
of supertubes of arbitrary shapes, this information is not enough to
fully determine the solution, however for the $U(1)\times U(1)$
invariant supertubes one can completely characterize the resulting
geometry.

First, the solution after the transition will still have the
$U(1)\times U(1)$ symmetry.  One can now use a theorem\foot{We thank
Harvey Reall for mentioning this paper to us.} that states \GibbonsSP\
that if a four-dimensional hyper-K\"ahler manifold has a $U(1)\times
U(1)$ symmetry then a linear combination of the two $U(1)$'s must be
tri-holomorphic\foot{Tri-holomorphic means that the $U(1)$ preserves
all three complex structures of the hyper-K\"ahler base.}  and hence
the metric must have Gibbons-Hawking form.  After the geometric transition,
the solution has two independent two-cycles and so the Gibbons-Hawking
space must have three centers.  The base space before and after the
transition must be asymptotic to $\IR^4$ and this means that the sum
of the GH charges at the three centers must be equal to one.  In order
to avoid singularities at the GH centers, the GH charges must be
integers, and so one center must have a negative charge.  Moreover, in
the limit when the dipole charges are small the solution must approach
the supertube in a flat base; hence the center at the origin of the
coordinate system must have charge $1$. The other two centers have
therefore charges $Q$ and $-Q$.  In this limit we expect these centers
to be located very close to each other, near the position of the ring
in the naive supertube geometry. Furthermore, the three centers must
also be colinear in order to preserve the $U(1)\times U(1)$ symmetry.

Thus, by using just a few facts about geometric transitions (which
came by a trivial extension of the physics seen in \LinNB) and the
fact that four-dimensional hyper-K\"ahler manifolds with $U(1)\times
U(1)$ symmetry are Gibbons-Hawking \GibbonsSP, we have reached the
conclusion that the singularity of the zero-entropy black ring is
resolved by the nucleation of a pair of oppositely charged
Gibbons-Hawking centers at the location of the ring.

One can now extend this argument to several concentric rings, and
observe that if one ring is resolved by the nucleation of one pair of
centers, several rings should be resolved by the nucleation of several
pairs of such points.  If the rings are concentric then the GH centers
should also be colinear.  One also expects that a solution with
several GH centers that are not colinear should be a simple
deformation of a bubbling supertube, and so it should also be dual to
a CFT microstate.  In this way one could expect the solutions to
contain pairs of equal but opposite GH charges; however, it is also
possible to deform such solutions so as to separate or combine the GH
centers.  Thus, the class of solutions that have a GH base and are
physically interesting should have any collection of GH centers of
positive and negative (integer) charges at arbitrary positions inside
the $\IR^3$ base of the Gibbons-Hawking space, with the constraint
that the GH charges sum to one.  More generally, if there is no
symmetry one should expect a general hyper-K\"ahler metric in which
the metric can flip from positive definite to negative definite.

We should also note that we expect these general multi-center
solutions to be bound states. One way to see this is to consider an
$n$-tube solution, which after the transition has $(2n+1)$ GH centers,
$(n+1)$ of which have positive charge.  Having resolved the geometry,
and perhaps separated GH charges still further, there is no canonical
way to pair up GH points and decide which pair forms a particular
tube. Moreover, in the limit when all the positively charged centers
coincide and all the negatively charged centers coincide, this
reproduces the bound state geometries of
\refs{\GiustoID,\GiustoIP,\GiustoKJ}. Another way to see that the 
multi-center geometries are bound states comes from the fact that one
cannot generically separate the centers into separated clusters
because of the fluxes wrapping the nontrivial $S^2$'s of the base.

In the next two sections we analyze this general solution. In Section
5 we construct the solution outlined above for the single bubbling
supertube and then we put this solution in a Taub-NUT background and
show that the singularity resolution mechanism derived in this section
reproduces the one found in the case of the zero entropy
four-dimensional black hole \BenaAY.

\newsec{Three-charge solutions with a Gibbons-Hawking base}

\subsec{The solutions in terms of harmonic functions}

In the M-theory frame, a background that preserves the same
supersymmetries as three sets of M2-branes can be written as
\refs{\BenaDE,\AdSBH}
\eqn\fullmet{\eqalign{ ds_{11}^2& =  - \left({1 \over Z_1 Z_2
Z_3}\right)^{2/3} (dt+k)^2 + \left( Z_1 Z_2 Z_3\right)^{1/3}
h_{mn}dx^m dx^n \cr &+ \left({Z_2 Z_3 \over
Z_1^2}\right)^{1/3}(dx_1^2+dx_2^2) + \left({Z_1 Z_3 \over
Z_2^2}\right)^{1/3}(dx_3^2+dx_4^2) + \left({Z_1 Z_2 \over
Z_3^2}\right)^{1/3}(dx_5^2+dx_6^2) \,,}}
\eqn\Aansatz{
 {\cal A}   ~=~  A^{(1)} \wedge dx_1 \wedge dx_2 ~+~  A^{(2)}   \wedge
dx_3 
 \wedge dx_4 ~+~ A^{(3)}  \wedge dx_5 \wedge dx_6\,,}
where $A^{(I)} $ and $k$ are one-forms in the five-dimensional space
transverse to the $T^6$. The metric, $h_{mn}$, is four-dimensional 
and hyper-K\"ahler.

When written in terms of the ``dipole field strengths''    $\Theta^I$,
\eqn\Thetadefn{\Theta^{(I)} \equiv d A^{(I)} + d\Big(  {dt +k \over
Z_I}\Big)} 
the BPS equations simplify to \refs{\BenaDE,\AdSBH}:
\eqn\eom{\eqalign{ \Theta^{(I)}  &= \star_4 \Theta^{(I)} \cr
\nabla^2  Z_I & = {1 \over 2  }  C_{IJK} \star_4 (\Theta^{(J)} \wedge
\Theta^{(K)}) \cr
dk + \star_4 dk &= Z_I \,  \Theta^{(I)} \,.}}
where $\star_4$ is the Hodge dual taken with respect to the
four-dimensional 
metric $h_{mn}$, and and $C_{IJK} ~\equiv~  |\epsilon_{IJK}|$. If the
$T^6$ is replaced by a more general Calabi-Yau manifold, the $C_{IJK}$
change accordingly.

We will take the base to have a Gibbons-Hawking metric:
\eqn\GHmet{h_{mn}dx^m dx^n  =V\, (dx^2 +  dy ^2 +   dz^2)~+~
{1 \over V}\, \big( d\psi + \vec{A} \cdot d\vec{y}\big)^2 \,,}
where we write $\vec y =(x,y,z)$ and where
\eqn\AVreln{\vec \nabla \times \vec A ~=~ \vec \nabla V\,.}

The solutions of \eom\ with a Gibbons-Hawking base have been derived
before in
\refs{\GauntlettQY,\GauntlettNW}. Here we derive them again using the
linear algorithm outlined in \BenaDE\  because we need some of the 
intermediate results. We consider a completely general base with
an arbitrary harmonic function, $V$.  We will denote the one-form,
$ \vec{A} \cdot d\vec{y} \equiv A$.  One should also recall that 
the coordinate $\psi$ has the range $0 \le \psi \le 4\,\pi$.

This metric has a natural set of frames:
\eqn\sdframes{\hat e^1~=~ V^{-{1\over 2}}\, (d\psi ~+~ A) \,,
\qquad \hat e^{a+1} ~=~ V^{1\over 2}\, dy^a \,, \quad a=1,2,3 \,.}
There are also two natural  sets of two-forms:
\eqn\sdforms{\Omega_\pm^{(a)} ~\equiv~ \hat e^1  \wedge \hat
e^{a+1} ~\pm~ \coeff{1}{2}\, \epsilon_{abc}\,\hat e^{b+1}  \wedge
\hat e^{c+1} \,, \qquad a =1,2,3\,.}
The $\Omega_-^{(a)}$ are anti-self-dual and harmonic, defining the
hyper-K\"ahler  structure on the base.  The forms, $\Omega_+^{(a)}$, are
self-dual, and we can take the  self-dual field strengths,
$\Theta^{(I)}$, to be proportional to them:
\eqn\thetaansatz{\Theta^{(I)} ~=~ - \, \sum_{a=1}^3 \,
\big(\partial_a \big( V^{-1}\, K^I \big)\big) \, \Omega_+^{(a)}
\,.}
For $\Theta^{(I)}$ to be closed, the functions $K^I$ have to be
harmonic in $\IR^3$.   Potentials satisfying $\Theta^{(I)} = dB^I$ are
then:
\eqn\Thetapots{ B^I ~\equiv~  V^{-1}\,   K^I   \,
(d\psi ~+~ A) ~+~ \vec{\xi}^I \cdot d \vec y \,,}
where
\eqn\xidefn{ \vec  \nabla \times \vec \xi^I ~=~ - \vec \nabla K^I\,.}
Hence, $\vec \xi^{I}$ are vector potentials for magnetic monopoles
located at the poles of $K^I$.

The three self-dual Maxwell fields $\Theta^{(I)}$ are thus determined
by the three harmonic functions $K^I$.   Inserting this result in
the right hand side of \eom\ we find:
\eqn\zzres{Z_I ~=~ \half \, C_{IJK} \, V^{-1}\,K^J K^K  ~+~ L_I \,,}
where $L_I$ are three more independent harmonic functions.

We now write the one-form, $k$, as:
\eqn\kansatz{k ~=~ \mu\, ( d\psi + A   ) ~+~ \omega}
and then the last equation in \eom\ becomes:
\eqn\roteqn{ \vec \nabla \times \vec \omega ~=~  ( V \vec \nabla \mu ~-~
\mu \vec \nabla V ) ~-~ \, V\, \sum_{i=1}^3 \,
 Z_I \, \vec \nabla \bigg({K^I \over V}\bigg) \,.}
Taking the divergence yields the following equation for $\mu$:
\eqn\mueqn{   \nabla^2 \mu ~=~ 2\, V^{-1}\, \vec \nabla \cdot
\bigg( V \sum_{i=1}^3 \, Z_I ~\vec \nabla {K^I \over V} \bigg)
\,,}
which is solved by:
\eqn\mures{\mu ~=~ \coeff {1}{6} \, C_{IJK}\,  {K^I K^J K^K \over V^2} ~+~
{1 \over 2 \,V} \, K^I L_I ~+~  M\,,}
where $M$ is yet another harmonic function.  Indeed, $M$
determines the anti-self-dual part of $dk$ that cancels out of the
last equation in \eom. Substituting this result for $\mu$ into
\roteqn\ we find that $\omega$ satisfies
\eqn\omegeqn{\vec \nabla \times \vec \omega ~=~  V \vec \nabla M ~-~
M \vec \nabla V ~+~   \coeff{1}{2}\, (K^I  \vec\nabla L_I - L_I \vec
\nabla K^I )\,.}

The solution is therefore characterized by the eight harmonic functions
$K^I, L_I$, $V$ and $M$.  Moreover, as observed in \BenaNI, the 
solutions are invariant under the shifts:
\eqn\gauge{\eqalign{K^I &~\ra~   K^I ~+~  c^I\, V \,,\cr L_I&~\ra~ L_I ~-~
C_{IJK}\,c^J \,K^K ~-~ \half C_{IJK}\, c^J\, c^K\, V \,,\cr M
&~\ra~ M -\half \,c^I \, L_I +{1 \over 12}\, C_{IJK}\left( V \, c^I \, c^J
\, c^K
+3\,  c^I\, c^J\, K^K\right) \,, }}
where the $c^I$ are three arbitrary constants. 

The eight functions that give the solution may be identified 
with the eight independent parameters that make up the $E_{7(7)}$ 
invariant as follows:
\eqn\Esevenreln{\eqalign{  x_{12} ~=~ & L_1 \,, \qquad x_{34} ~=~ L_2 \,, 
\qquad x_{56} ~=~ L_3  \,, \qquad x_{78} ~=~ - V  \,,  \cr 
y_{12} ~=~ & K^1 \,, \qquad y_{34} ~=~ K^2 \,, 
\qquad y_{56} ~=~ K^3  \,, \qquad y_{78} ~=~ 2\, M \,.}}
With these identifications, one can identify the right-hand side of
\omegeqn\  in terms of the symplectic invariant of the {\bf 56} of
$E_{7(7)}$:
\eqn\newomega{\vec \nabla \times \vec \omega ~=~   \coeff{1}{4}\, 
\sum_{A,B} \, (y_{AB}  \vec\nabla x_{AB} ~-~ x_{AB}  \vec\nabla y_{AB}
)\,.}

For future  reference, we note that the quartic invariant of  the {\bf 56}
of $E_{7(7)}$
is determined by:
\eqn\quarticE{\eqalign{ J_4 ~=~ & -{1 \over 4}(x_{12}y^{12} + x_{34}y^{34}
+x_{56}y^{56}+x_{78}y^{78})^2-
(x_{12}x_{34}x_{56}x_{78}+y^{12}y^{34}y^{56}y^{78})\cr & +
x_{12}x_{34}y^{12}y^{34}+ x_{12}x_{56}y^{12}y^{56} +
x_{34}x_{56}y^{34}y^{56}+x_{12}x_{78}y^{12}y^{78}+ x_{34}x_{78}
y^{34}y^{78}\cr &+x_{56}x_{78}y^{56}y^{78} \,.}}
%

\subsec{Dirac-Misner strings and closed time-like curves}

To look for the presence of closed time-like curves in the metric one 
considers  the space-space components of the metric given by \fullmet\ and 
\GHmet\  in the direction of the base.  If we denote $W \equiv (Z_1 Z_2
Z_3)^{1/6}$,
and use  the expression for $k$ in \kansatz\ then we find
\eqn\dstil{\eqalign{ ds^2_4   ~=~ & - W^{-4}\, \big( \mu  (d \psi+ A ) +
\omega  \big)^2 \cr & \qquad ~+~ {W^2 V^{-1}}\big(
d\psi + A \big)^2 + W^2 V \big(dr^2 + r^2 d\theta^2 + r^2
\sin^2 \theta \, d\phi^2\big) \cr
 ~=~ & W^{-4}\,  (W^6 V^{-1}-\mu^2 )\Big(  d\psi + A  -
{\mu \,  \omega  \over W^6 V^{-1}-\mu^2 }  \Big)^2  -
 {W^2 \, V^{-1}   \over W^6 V^{-1}-\mu^2 } \,  \omega^2\cr
& ~+~  W^2 V \big(dr^2 + r^2 d\theta^2 + r^2
\sin^2 \theta\, d\phi^2\big) \cr ~=~ & {\cQ \over W^4 V^2} \Big(
d\psi + A  - {\mu \, V^2 \over \cQ }\, \omega \Big)^2 +
W^2 V \Big( r^2 \sin^2 \theta \, d \phi^2 -{\omega^2  \over \cQ} \Big)
+ W^2 V (dr^2 + r^2 d\theta^2) \,, }}
where we have introduced the quantity: 
\eqn\Qdefn{\cQ ~\equiv~  W^6\, V ~-~  \mu^2\,  V^2 ~=~  Z_1 Z_2 Z_3 V ~-~
\mu^2 \, V^2\,.}
We have also chosen  to write the metric on $\IR^3$ in terms of a generic
set of  spherical polar coordinates,  $(r,\theta, \phi)$.

Upon evaluating $\cQ$ as a function of the eight
harmonic functions that determine the solution one obtains a
beautiful result:
\eqn\qqq{\eqalign{ \cQ &= - M^2\,V^2   - {1 \over
3}\,M\,C_{IJK}{K^I}\,{K^J}\,{K^k} - M\,V\,{K^I}\,{L_I} - {1 \over
4}(K^I L_I)^2\cr &\quad+{1 \over 6} V C^{IJK}L_I L_J L_K +{1 \over
4} C^{IJK}C_{IMN}L_J L_K K^M K^N}}
with $C^{IJK}=C_{IJK}$.  We see that  $\cQ$  is nothing other than the
$E_{7(7)}$ quartic invariant \quarticE\ where the $x$'s and 
$y$'s are identified as in \Esevenreln.

From \dstil\ and \fullmet\ we see that to avoid CTC's, the following 
inequalities must be true everywhere:
\eqn\Qpos{\cQ ~ \ge ~ 0\,, \qquad W^2 \, V ~ \ge ~ 0\,, \qquad 
\big(Z_J \, Z_K \, Z_I^{-2}\big)^{1\over3}
~=~ W^2 Z_I^{-1}~ \ge ~ 0\,, \ \ I =1,2,3 \,.}

The last two conditions can be subsumed into:
\eqn\VZpos{V\, Z_I ~=~  \coeff {1}{2}\, C_{IJK}\,  K^J \, K^K ~+~ L_I\,  V~ \ge ~ 0 \,,
\qquad  I = 1,2,3\,.}
The obvious danger arises when $V$ is negative.   We will show in the next 
sub-section that all these quantities remain finite and positive in a 
neighborhood  of $V=0$, despite the fact that $W$ blows up. Nevertheless, 
these quantities could possibly be negative away from the $V=0$ surface. 
While we will, by no means, make a complete analysis of the positivity
of these quantities, we will discuss it further in section 4, and show
that \VZpos\ does not present a significant problem in a simple example.
One should also note that $\cQ \ge 0$ requires $\prod_I (V Z_I) \ge
\mu^2 V^4$, and so generically the constraint  $\cQ \ge 0$ 
is stronger then the constraints \VZpos.\foot{There might of course exist 
some solutions where two of the  $ V Z_I$ change sign on exactly the same 
codimension one surface, but these are non-generic.} 

Having imposed these conditions there is evidently another potentially
dangerous term: $r^2 \sin^2 \theta \, d \phi^2 -{\omega^2
\over \cQ}$.  There will be CTC's if the first term does not always dominate 
over the second.  In particular, one will have CTC's if $\omega$
remains finite as one moves onto the polar axis where $\theta =0,
\pi$.  This happens precisely when there is a Dirac-Misner
string\foot{In terms of vector fields, Dirac strings and Dirac-Misner
strings are the same thing, but we use the former term for vector
potentials of Maxwell fields, and we use the latter when the vector is
part of the metric. The latter is a potentially dangerous physical
singularity, unless it can be unwound by a non-trivial $U(1)$
fibration.}  in the metric.  Thus to avoid CTC's we must make sure
that the solution has no Dirac-Misner strings anywhere.

\subsec{Ergospheres}

As we have seen, the general solutions we will consider have 
functions, $V$, that
change sign on the $\IR^3$ base of the GH metric.  Our purpose here is to show
that such solutions are completely regular, with positive definite
metrics, in the regions where $V$ changes sign.  As we will see the 
surfaces $V=0$ amount to a set of completely harmless ergospheres.

The most obvious issue is that if $V$ changes sign, then the overall 
sign of the metric \GHmet\ changes and there might be a large number
of closed time-like curves when $V<0$.    However, we remarked above that
the warp factors, in the form of $W$, prevent this from
happening.  Specifically,
the expanded form of the  complete, eleven-dimensional metric when
projected 
onto the GH base  yields \dstil. In particular, one has
\eqn\WVsq{W ^2 \, V ~=~ (Z_1\, Z_2 \, Z_3 \, V^3)^{1\over 3} ~\sim~ 
((K_1\, K_2 \, K_3)^2)^{1\over 3} }
on the surface $V=0$.  Therefore, $W ^2  V$ is regular and positive on 
this surface. 

There is still the danger of singularities at $V=0$ for the other
background 
fields.
We first note that there is no danger of such singularities being hidden
implicitly in
the $\vec \omega$ terms.  Even though \roteqn\ suggests that the
source of $\vec \omega$ is singular at $V=0$, we see from \omegeqn\ that
the source is regular at $V=0$ and thus there is nothing hidden in $\vec
\omega$.
We therefore need to focus on the explicit inverse powers of $V$ in the
solution.

First, the factors of $V$ cancel in the torus warp factors, which are
of the form $(Z_I Z_j Z_K^{-2})^{1 \over 3}$. The coefficient of $(dt
+ k)^2$ is $W^{-4}$, which vanishes as $V^2$. The singular part of the
cross term, $ dt \, k$, is the $\mu \, dt \,(d \psi +A)$, which, from
\mures, diverges as $V^{-2}$, and so the cross term remains finite at
$V=0$.  So the metric, and the spatial parts of the inverse metric are
regular at $V=0$.  This surface is therefore not an event horizon. 
 It is, however, a Killing
horizon or, more specifically, an ergosphere: The time-like Killing
vector defined by translations in $t$ becomes null when $V=0$.

At first sight, it does appear that the Maxwell fields are singular on the
surface
$V=0$.  Certainly the ``magnetic components,'' $\Theta^I$, in
\thetaansatz\ 
are singular when $V=0$.  However, one knows that the metric is
non-singular
and so one should expect that the singularity in the $\Theta^I$ to be
unphysical.
This intuition is correct:  One must remember that the complete Maxwell
fields
are the $A^{(I)}$, and these are indeed non-singular at $V=0$.  One finds
that the singularities in the ``magnetic terms'' of $A^{(I)}$ are canceled
by 
singularities in the ``electric terms'' of $A^{(I)}$, and this is possible
at $V=0$
precisely because it is an ergosphere and the magnetic and electric terms
can communicate.  Specifically, one has, from \Thetadefn\ and \Thetapots:
\eqn\dAI{d A^{(I)}~=~ d\, \bigg(B^{(I)} ~-~ {(dt + k) \over
Z_I}\bigg) \,.}
Near $V=0$ the singular parts of this behave as:
\eqn\dAI{\eqalign{d A^{(I)}~\sim~ & d\, \bigg({ K^I \over V} ~-~ { \mu
\over Z_I}\bigg) \,
(d \psi +A) \cr ~\sim~  & d\, \bigg({ K^I \over V} ~-~ { K^1 \,  K^2 \,K^3
\over 
\half V\, C_{IJK}\,   K^J \,K^K}\bigg) \, (d \psi +A)~\sim~0\,.}}
The cancellations of the $V^{-1}$ terms here occur for much the same
reason
that they do in the metric \dstil.

Therefore, even if $V$ vanishes and changes sign and the base metric
becomes negative definite, the complete eleven-dimensional solution is
regular and well-behaved around the $V=0$ surfaces.  
It is this fact that gets us around the uniqueness
theorems for asymptotically Euclidean self-dual (hyper-K\"ahler) metrics
in four dimensions, and as we will see, there are now a vast number of
candidates for the base metric.

\newsec{Constructing explicit solutions}

\subsec{The harmonic functions}

We now specify the type of harmonic functions that will underlie our
solutions.  In particular, we will consider functions with a finite
set of isolated sources. 
Let $\vec{y}^{(j)}$ be  the positions of the source points in the $\IR^3$
of the 
base, and let  $r_j \equiv |\vec{y}-\vec{y}^{(j)}|$.  We take:
\eqn\Vform{ V = \varepsilon_0 ~+~ \sum_{j=1}^N \,  {q_j  \over r_j} \,,} 
where $\varepsilon_0$ can be chosen to be $1$ if the base is
asymptotically Taub-NUT and $\varepsilon_0 = 0$ for an asymptotically
Euclidean (AE) space.  If $q_j \in \ZZ$ then the metric at $r_j =0$
has a (spatial) $\ZZ_{|q_j|}$ orbifold singularity, but this is benign
in string theory, and so we will view such backgrounds as regular.  It
is convenient to define:
\eqn\qzerodefn{q_0 ~\equiv~ \sum_{j=1}^N \, q_j \,,}
and note that the metric is asymptotic to $\IR^4$ if and only if
$|q_0| =1$.  By convention we will take $q_0 >0$.  This means that $V$
is positive for large $r$.  Moreover the fact that $q_j \in \ZZ$ means
that the only non-trivial backgrounds will have some negative $q_j$'s,
and thus the function $V$ will be negative in the vicinity of these GH
centers.  As we saw in section 3, the surfaces $V=0$ when $V$ changes
sign.

We can  choose the harmonic functions $K^I,L_I$ and $M$ to be 
localized anywhere on the base. These solutions have localized brane
sources, and include for example the supertube and the 
black ring in Taub-NUT \refs{\BenaAY,\GaiottoXT,\ElvangSA,\BenaNI}.
However, as explained in Section 2, we are interested in the solutions 
without localized branes, so we consider harmonic 
functions $K^I, L^I$ and $M$ 
whose singularities are localized at the GH centers:
\eqn\KLMexp{K^I ~=~\kch^I_0 ~+~  \sum_{j=1}^N \, {\kch_j^I \over r_j} \,,
\quad 
L^I ~=~ \lch^I_0 ~+~  \sum_{j=1}^N \, {\lch_j^I \over r_j} \,, \quad 
M ~=~ \mch_0 ~+~  \sum_{j=1}^N \, {\mch_j \over r_j} \,.}
%
\subsec{Cycles and fluxes}

The multiple-center GH base \Vform\ has many non-contractible two-cycles, 
$\Delta_{ij}$, that run between the GH centers. 
These two-cycles can be defined by 
taking any  curve, $\gamma_{ij}$, between $\vec{y}^{(i)}$  and
$\vec{y}^{(j)}$ and 
considering the $U(1)$ fiber of \GHmet\ along the curve.  This fiber
collapses  to  zero at the GH centers, and so the curve and the fiber
sweep out a $2$-sphere (up to $\ZZ_{|q_j|}$ orbifolds).  There are 
$(N-1)$ linearly independent homology
two-spheres, and  the set $\Delta_{i\, (i+1)}$ represents a basis.  These
spheres intersect one another at the points $\vec{y}^{(k)}$.

The fluxes that thread these two-cycles depend on the behavior of the
functions, $K^I$ at the GH centers. To determine the fluxes we
need the explicit forms for the vector potentials, $B^I$, in
\Thetapots, and to find these we first need the vector fields, $\vec v_i$,
that satisfy:
\eqn\vieqn{  \vec \nabla \times \vec v_{i} ~=~    
\vec \nabla\, \bigg( {1\over r_i} \bigg) \,.}
One then has: 
\eqn\Asoln{\vec A ~=~  \sum_{j=1}^N \,  q_j \, \vec v_j \,, \qquad 
\vec \xi^I ~=~  \sum_{j=1}^N \,  \kch_j^I \, \vec v_j \,.}
If we choose coordinates so that $\vec y^{(i)} ~=~ (0,0,a)$, 
 $(y_1,y_2, y_3) = (x,y,z)$  and let $\phi$ denote the polar angle in the 
$(x,y)$-plane, then:
\eqn\vzdefns{ \vec v_{i} \cdot d \vec y~=~ \Big( {(z -a) \over r_i} ~+~
c_i \Big) \, d \phi \,,}
where $c_i$ is a constant.  The  vector field, $\vec v_i$, is regular away 
from the $z$-axis, but has a Dirac string along the $z$-axis.  By choosing
$c_i$ we can cancel the string along the positive or negative
$z$-axis, and by moving the axis we can arrange these strings  to run in
any  direction we choose, but they must start or finish at some $\vec y^{(i)}$,
or run out to infinity.

Now consider what happens to $B^I$ in the neighborhood of
$\vec y^{(i)}$.  Since the circles swept out by $\psi$ and $\phi$ 
are shrinking to zero size, 
the string singularities near $\vec y^{(i)}$ are of the form:
\eqn\BIasymp{B^I ~\sim~ { \kch_i^I \over q_i} \, \Big (d \psi ~+~ 
q_i \,\Big( {(z -a) \over r_i} ~+~ c_i \Big) \, d \phi \Big)  ~-~ 
\kch_i^I \, \Big( { (z -a) \over r_i} ~+~ c_i \Big) \, d \phi   ~\sim~
 { \kch_i^I \over q_i} \, d \psi \,.}
This shows that the vector, $\vec \xi^I$, in \Thetapots\  cancels the
string  singularities in the $\IR^3$ for each of the complete vector fields,
$B^I$.   The singular components of $B^I$ thus point along the 
$U(1)$ fiber of the GH metric. 

If we choose any curve, $\gamma_{ij}$,  between $\vec{y}^{(i)}$  and 
$\vec{y}^{(j)}$ then the vector fields, $B^I$, are regular over the 
whole $\Delta_{ij}$ except at the end-points, $\vec{y}^{(i)}$  
and  $\vec{y}^{(j)}$.  Let $\widehat \Delta_{ij}$ be the 
cycle $\Delta_{ij}$ with the poles excised.
Since   $\Theta^{(I)}$ is regular at the poles, we have
\eqn\periods{\eqalign{\flux^{(I)}_{ij} ~\equiv~ & {1 \over 4\, \pi}\,
\int_{\Delta_{ij}} \, 
\Theta^{(I)} ~=~ {1 \over 4\, \pi}\, \int_{\widehat \Delta_{ij}} \,
\Theta^{(I)} ~=~ 
{1 \over 4\, \pi}\, \int_{\del \widehat  \Delta_{ij}} \, B^{(I)}  \cr  ~=~& 
{1 \over 4\, \pi}\,  \int_0^{4\pi} \,  d \psi \, \big( B^{(I)}|_{y^{(j)}}
~-~
 B^{(I)}|_{y^{(i)}}  \big) ~=~   \bigg( { \kch_j^I \over q_j} ~-~  
 { \kch_i^I \over q_i} \bigg) \,.}}
We have normalized these periods for later convenience, and they give 
the $I^{\rm th}$-flux threading the cycle $\Delta_{ij}$.  As we will see, these
fluxes are directly responsible  for holding up the cycle.

\subsec{Solving for $\omega$}

Since everything is determined by the eight harmonic functions \KLMexp, all
that remains is to solve for $\omega$ in equation \omegeqn.  The right-hand 
side of \omegeqn\ has two kinds of terms:
\eqn\rhsterms{{1\over r_i}\, \vec \nabla\, {1\over r_j} ~-~ {1\over r_j}\, \vec \nabla\,
{1\over r_i}   \qquad {\rm and } \qquad \vec \nabla \, {1\over r_i} \,.}
Hence  $\omega$  will be built from the vectors $\vec v_{i}$ of 
\vieqn\ and some new vectors,$ \vec w_{ij}$, defined by:
\eqn\wijeqn{ \vec \nabla \times \vec w_{ij} ~=~  {1\over r_i}\, \vec
\nabla\, {1\over r_j} ~-~ {1\over r_j}\, \vec \nabla\, {1\over r_i} \,.}

To find a simple expression for $ \vec w_{ij}$ it is convenient to
use the coordinates outlined above with the $z$-axis running
through $\vec y^{(i)}$ and  $\vec y^{(j)}$. Indeed, 
choose coordinates so that $\vec y^{(i)} ~=~ (0,0,a)$ and $\vec
y^{(j)} ~=~ (0,0,b)$ and one may take $a > b$.   Then the explicit solutions may be 
written very simply: 
\eqn\wijres{w_{ij} ~=~ - {(x^2 +  y^2 + (z-a)(z-b)) \over (a-b)  \, r_i  \, r_j }
\, d \phi \,.}
This is then easy to convert  to a more general system of coordinates.  One
can then add up all the contributions to $\omega$ from all the
pairs of points.

There is, however, a more convenient basis of vector fields that
may be used instead of the $w_{ij}$.  Define:
\eqn\omegaijdefn{ \omega_{ij} ~\equiv~ w_{ij} + 
\coeff{1}{(a-b)} \, \big(  v_{i} - v_{j}  + d\phi \big) ~=~ - 
{(x^2 +  y^2 + (z-a+ r_i)(z-b - r_j)) \over (a-b)  \, r_i  \, r_j } \, d \phi  \,,}
These vector fields then satisfy:
\eqn\omijeqn{  \vec \nabla \times \vec \omega_{ij} ~=~  {1\over r_i}\, \vec
\nabla\, {1\over r_j} ~-~ {1\over r_j}\, \vec \nabla\, {1\over r_i}~+~
 {1 \over \dist_{ij} } \, \bigg(\vec \nabla\,  {1\over r_i} ~-~ \vec \nabla\,  
 {1\over r_j} \bigg) \,,}
where 
\eqn\dijdefn{\dist_{ij} ~\equiv~ |\vec y^{(i)} ~-~ \vec y^{(j)}  | }
is the distance between the $i^{\rm th}$ and $j^{\rm th}$ center
in the Gibbons-Hawking metric.

We then see that the general solution for $\vec \omega$ may be written
as:
\eqn\omsol{\vec \omega ~=~  \sum_{i,j}^ N \, a_{ij} \, \vec \omega_{ij} ~+~ 
 \sum_{i}^ N \, b_{i} \, \vec v_{i}\,,}
for some constants $a_{ij}$, $b_i$.

The important point about the $\omega_{ij}$ is that they have {\it no
string singularities whatsoever}, and thus they can be used to solve
\omegeqn\ with the first set of source terms in \rhsterms, without
introducing Dirac-Misner strings, but at the cost of adding new source
terms of the form of the second term in \rhsterms.  If there are $N$
source points, $\vec{y}^{(j)}$, then using the $w_{ij}$ suggests that
there are ${1 \over 2} N(N-1)$ possible string singularities
associated with the axes between every pair of points $\vec{y}^{(i)}$
and $\vec{y}^{(j)}$.  However, using the $\omega_{ij}$ makes it far
more transparent that all the string singularities can be reduced to
those associated with the second set of terms in \rhsterms\ and so
there are at most $N$ possible string singularities and these can be
arranged to run in any direction from each of the points
$\vec{y}^{(j)}$.

However, for non-singular solutions and, as we have seen, to avoid CTC's,
we must find solutions without  Dirac-Misner strings.  
The vector potentials, $\vec v_i$,  necessarily have 
such singularities, and therefore string singularities will arise
through the second term in \omsol.  These strings originate from 
each $\vec y^{(j)}$, and while they can be arranged to coincide and cancel in 
some places, there will always be regions that have non-trivial strings.  
We therefore have to require that the solution for $\omega$
be constructed entirely out of the $\omega_{ij}$ in \omijeqn.  That is,
we must require that $b_i =0$ in \omsol.
This yields a set of $N$ constraints that relate the charges  and distances,
$\dist_{ij}$.   We will refer to these as the ``bubble equations.''

\subsec{The non-singular solutions}
 
We have seen that the constants $q_j$ and $\kch_i^I$ determine the
geometry and the fluxes in the solution.  We now fix the remaining
constants, $\lch_i^I $ and $\mch_i$, by requiring that the solutions
have no sources for the brane charge. With but a few exceptions, non-zero 
sources for the brane charge will lead to singularities or black hole
horizons, and are better avoided if one wants to construct microstate solutions.

As one approaches $r_j =0$ one finds:
\eqn\Zasymp{Z_I ~\sim ~ \bigg( \half \, C_{IJK} \,  { \kch_j^J \, \kch_j^K 
\over q_j}  ~+~ \lch_j^I  \bigg)\,  {1 \over r_j}  \,.}
We thus remove the brane sources by choosing:  
\eqn\lchchoice{ \lch_j^I ~=~ -  \half \, C_{IJK} \,  { \kch_j^J \, \kch_j^K 
\over q_j} \,, \quad j=1,\dots, N \,.}
Since there are no brane sources, $\mu$ cannot be allowed to diverge  at 
$r_j =0$, which determines: 
\eqn\mchchoice{
\mch_j ~=~  \coeff {1}{12} C_{IJK} {\kch_j^I \, \kch_j^J \, \kch_j^K \over q_j^2}    ~=~ 
\half\,  {\kch_j^1 \, \kch_j^2 \, \kch_j^3 \over q_j^2}    \,.}

The constant terms in \KLMexp\ determine the behavior of the solution
at infinity.  If the asymptotic geometry is Taub-NUT, all these term
can be nonzero, and they correspond to combinations of the moduli. (A
more thorough investigation of these parameters can be found in the
last section of \BenaNI.)  However, in order to obtain solutions that
are asymptotic to five-dimensional Minkowski space, $\IR^{4,1}$, one
must take $\varepsilon_0 = 0$ in \Vform, and $k_0^I =0$. Moreover,
$\mu $ must vanish at infinity, and this fixes $m_0$.  For simplicity
we also fix the asymptotic values of the moduli that give the size of
the three $T^2$'s, and take $Z_I \to 1$ as $r \to \infty$. Hence, the
solutions that are asymptotic to five-dimensional Minkowski space
have:
\eqn\fiveDsol{\varepsilon_0 = 0 \,, \qquad  k_0^I =0\,, \qquad   l_0^I =1\,, \qquad  
 m_0  = -\half \, q_0^{-1} \, \sum_{j=1}^N\, \sum_{I=3}^N k_j^I \,.}
It is straightforward to generalize our results to solutions with 
different asymptotics, and in particular to Taub-NUT.

As we observed above, we must find a solution with no Dirac-Misner strings, and
thus $\omega$ must be made out of the $\omega_{ij}$ in \omijeqn.  To match the 
$ {1\over r_i}\, \vec \nabla\, {1\over r_j} $ terms on the right hand side of 
\omegeqn\  we must take: 
\eqn\omegacomplete{\eqalign{\vec  \omega   ~\equiv~&  \half \, \sum_{i, j =1}^N \, 
\Big( \Big(\half  \,\sum_{I=1}^3  (k^I_i \, l^I_j ~-~  l^I_i \, k^I_j)   \Big)  ~+~ 
(q_i \, m_j ~-~  m_i \, q_j)  \Big) \, \vec \omega_{ij} \cr ~=~& \coeff{1}{4} \,
 \sum_{i, j =1}^N \, q_i \, q_j \, \bigg(\prod_{I=1}^3 \, \bigg(  {k^I_j \over q_j} ~-~  
 {k^I_i \over q_i}  \bigg)\bigg) \, \vec \omega_{ij} ~=~  \coeff{1}{4} \,
 \sum_{i, j =1}^N \, q_i \, q_j \,   \flux^{(1)}_{ij} \,  \flux^{(2)}_{ij} \,  \flux^{(3)}_{ij}
 \,  \vec \omega_{ij}  \,.}}
This then satisfies
\eqn\partsol{\eqalign{
\vec \nabla \times \vec   \omega    ~-~  & \big( V \vec \nabla M ~-~
M \vec \nabla V ~+~   \coeff{1}{2}\, (K^I  \vec\nabla L_I - L_I \vec
\nabla K^I ) \big)  \cr   ~=~ &   \sum_{i =1}^N \,  
\Big(m_0 \, q_i ~+~  \half   \sum_{I=1}^3  k^I_i  \Big)\,  \vec \nabla\, \bigg({1\over r_i} 
\bigg)  \cr   & \qquad  ~+~   \coeff{1}{4} \,  \sum_{i, j =1}^N \, q_i \, q_j \,   \flux^{(1)}_{ij} \, 
 \flux^{(2)}_{ij} \,  \flux^{(3)}_{ij} \,  
 {1 \over \dist_{ij} } \, \bigg(\vec \nabla\,  {1\over r_i} ~-~ \vec \nabla\,  
 {1\over r_j} \bigg)\,,}}
and the absence of CTC's means that the right-hand side of this must vanish.
Collecting terms in $ \vec \nabla (r_j^{-1})$ and requiring that each of them vanish
leads to a system of ``bubble equations,''  relating $r_{ij}$ to the fluxes:
\eqn\linsys{   \sum_{{\scriptstyle j=1} \atop {\scriptstyle j \ne i}}^N \, 
\,  \flux^{(1)}_{ij} \,   \flux^{(2)}_{ij} \,  \flux^{(3)}_{ij} \   {q_i \, q_j  \over \dist_{ij} } ~=~
-2\, \Big(m_0 \, q_i ~+~  \half   \sum_{I=1}^3  k^I_i \Big) \,.}

The solution for $\omega$ in \omegacomplete, and the equations
\linsys\ can be trivially extended to more complicated $U(1)^n$
supergravity theories by replacing $ \flux^{(1)}_{ij} \,
\flux^{(2)}_{ij} \, \flux^{(3)}_{ij} $ with $\coeff {1}{6} C_{IJK}
\flux^{(I)}_{ij} \, \flux^{(J)}_{ij} \, \flux^{(K)}_{ij} $, and 
replacing products of quantities like $\prod_{I=1}^3 X^I$ by ${1\over
6} C_{IJK} X^I X^J X^K$.

Note that if one sums \linsys\ over all values of $i$ then the 
skew symmetry of the left-hand side causes it to vanish.  The result
is:
\eqn\redund{  \sum_{i =1}^N \,   \Big(m_0 \, q_i ~+~  \half   \sum_{I=1}^3 
 k^I_i \Big) ~=~ 0\,,}
which is the last equation in \fiveDsol.  Thus there are generically only
$(N-1)$ independent ``bubble equations'' in \linsys.  

We see from \linsys\ that the $\dist_{ij}$ are related directly to the fluxes, but
for $N>2$, the $\dist_{ij}$ are not fixed by a choice of fluxes:  There are
moduli, and we will discuss this below.  Also  note that 
if  {\it any one} of the fluxes $\flux_{ij}^I$, $I=1,2,3$ vanishes, then
the $\dist_{ij}$ drops out of the equations completely. 
 
The other important constraint is \Qpos.  We are not going to make a complete
analysis of this, but we note that the only obvious danger points are when
$r_j =0$ for some $j$.   For the non-singular solutions considered here we have
$Z_i$ going to a constant at $r_j =0$, and from \Qdefn\ we see that $\cQ$ will 
become negative unless:
\eqn\mucond{ \mu(\vec y = \vec y^{(j)} ) ~=~ 0\,, \quad j=1,\dots,N \,.}
It turns out that this set of constraints is exactly the same as the set of
equations \linsys.  We have checked this explicitly, but it is also rather
easy to see from \roteqn.  The string singularities in $\vec \omega$ 
potentially arise from the $\vec \nabla (r_j^{-1})$ terms on
the right-hand side of \roteqn.  We have already arranged that the $Z_i$ 
and $\mu$ go to finite limits at $r_j =0$, and the same is automatically true
of $K^I V^{-1}$.  This means that the only term on the right hand side of
\roteqn\ that could, and indeed will, source a string is the $\mu \vec \nabla V$ 
term.  Thus removing the string singularities is equivalent to \mucond. 
Moreover the fact that the sum of the resulting equations reduces to \redund\
is simply because the condition $\mu \to 0$ as $r \to \infty$ means that there
is no Dirac-Misner string running out to infinity.

For the non-singular solutions it is easy to check that 
\eqn\VZIns{V\, Z_I ~=~ \sum_{i=1}^N \, {q_i \over r_i} ~-~ \coeff{1}{4}
\, C_{IJK}\,  \sum_{i, j =1}^N \, \Pi^J_{ij}  \, \Pi^K_{ij}  \, 
{q_i\, q_j \over r_i \, r_j} \,,}
and, as we noted in \VZpos, this must be positive.  While we have
not been able to show this is true in general, we suspect that the 
positivity of these functions will follow from  the bubble equations, \linsys, 
triangle inequalities between $r_i$, $r_j$ and $r_{ij}$, 
and some simple constraints on the charges.  We will consider a very
simple example below.

At this point  it is very instructive to count parameters. 
There are $3N$ charges, $k_j^I$,
and the set of points, $\vec y^{(j)}$, have $3N$ parameters. The Euclidean
$\IR^3$ of the base has three translational symmetries and three rotational
symmetries, which means that, for $N \ge 3$, the generic solution has $6(N-1)$ 
parameters.   The equations impose $(N-1)$ constraints, leaving 
$5(N-1)$ free parameters.    For $N=2$ there is a residual axi-symmetry 
which means that there are $5N-4 = 6$ parameters, which correspond to the
choice of the $k^I_j$. We should also remember that three combinations of the $k^I_j$
do not affect the final solution, because of the gauge invariance \gauge.

Physically, it is natural to fix the charges and solve \linsys\ to determine
some of the $\dist_{ij}$ in terms of other $\dist_{l m}$.  Note that doing this
turns \linsys\ into a linear system for the $\dist_{ij}^{-1}$, which is elementary
to invert.  In finding the solutions one must  remember to impose the 
triangle inequalities:
\eqn\triangles{\dist_{ij} ~+~ \dist_{jk} ~\ge~ \dist_{ik} \,, \quad \forall  i,j,k \,.}

It is easy to see that there is always a physical solution for some
ranges of the $\dist_{ij}$.  Indeed, if put all the points $\vec
y^{(j)}$ on a single axis, which means the complete solution preserves
a $U(1) \times U(1)$ symmetry overall, then the only geometric
parameters are the $(N-1)$ separations of the points on the axis.  The
triangle inequalities are all trivially satisfied.  Thus only $(N-1)$
of the $\dist_{ij}$ are independent, and they are uniquely
fixed\foot{If, in solving \linsys, one finds one of the independent
$\dist_{ij}$ to be negative, one can render it positive by reordering
the points.} by \linsys\ in terms of the choice of flux parameters,
$k^I_j$.  One can then vary the $\dist_{ij}$ about this solution and
one finds the complete family with all allowed ranges of $\dist_{ij}$
consistent with the triangle inequality.

It is rather easy to understand the physical picture of the solution
set.  The fluxes, determined by $k^I_j$, are holding up the blown up
cycles, whose sizes are determined by the $\dist_{ij}$.  If one puts
all the cycles in a straight line, then their sizes are fixed uniquely
by the magnitudes of the fluxes through the cycles. One can think of
these cycles as being characterized by ``rods'' of length $\dist_{i
\,(i+1)}$ along the axis. However, the cycles can move around, and so
the rods can pivot about their end-points while remaining connected to
one another.  The rod lengths can vary, but they generically vary only
a small amount: Their length is set by the fluxes through the cycles,
and these are modified only when neighboring cycles get close enough
that they interfere with each others fluxes.  Rods can, however
combine and break when a point $\vec y^{(\ell)}$ crosses the axis
$\vec y^{(i)}$ and $\vec y^{(j)}$.  Put more mathematically, by
imposing axi-symmetry one gets a preferred homology basis with fixed
scales.  This basis will generically undergo Weyl reflections when the
difference of two cycles collapses.

\subsec{Examples}

Consider $V$ with three charges:
\eqn\threeq{q_1 ~=~ 1\,, \qquad q_2 ~=~  1\,, \qquad q_3 ~=~ -1\,.}
Since we have $q_0 \equiv \sum q_j =1$,  the base is asymptotic to $\IR^4$.
The equations \linsys\ reduce to a system of the form:
\eqn\threechsys{
 -  {A_{12} \over \dist_{12}} ~+~ {A_{13}  \over \dist_{13}}  ~=~    (B_2+ B_3)\,,  \quad 
 {A_{12} \over \dist_{12}} ~+~ {A_{23}  \over \dist_{23}}    ~=~    (B_1 + B_3)\,, \quad
 {A_{13} \over \dist_{13}} ~+~  {A_{23}  \over \dist_{23}} ~=~  (B_1 + B_2 + 2\, B_3)  \,.}
where 
\eqn\ABforms{\eqalign{A_{12} ~\equiv~ & \prod_{I=1}^3 \, ( k^I_1 - k^I_2) \,, \quad   
A_{13} ~\equiv~ \prod_{I=1}^3 \, ( k^I_1 +  k^I_3) \,, \quad A_{23} ~\equiv~ 
\prod_{I=1}^3 \,  ( k^I_2 + k^I_3) \,,  \cr B_{j} ~\equiv~&  \sum_{I=1}^3 \, k^I_j\,, 
\quad j=1,2,3   \,.}}
If we impose the condition that the Gibbons-Hawking centers are co-linear,
for example $\dist_{13} =\dist_{12}+\dist_{23}$, then the forgoing equations
reduce to a quadratic equation.  The ordinary, zero-entropy
black ring emerges as $\dist_{23} \to 0$,
and so we know there is certainly a family of solutions in this limit.  

Now suppose we have a solution, and we want to find the family to
which it belongs.  We only need use two of the equations in
\threechsys, and so consider the first two equations.  Choose
$\dist_{12}$ to have the value for the known solution.  The first two
equations in \threechsys\ tell us that the third charge ($j=3$) must
be located at a determined distance from each of the other two
charges, {\it i.e.}  $\dist_{13}$ and $\dist_{23}$ are fixed.  This
determines the location of the third charge up to rotations about the
axis through the first two charges.  Suppose we now decrease
$\dist_{12}$ by a small amount. The first two equations in
\threechsys\ tell us that if $\pm {A_{13} \over A_{12}} > 0$ then $\pm
\dist_{13}$ must decrease and if $\pm {A_{23} \over A_{12}} > 0$ then
$\pm \dist_{23}$ must increase.  Thus the third charge must move
around to compensate, and if $\dist_{13}$ and $\dist_{23}$ change in
opposite senses then the third charge will move in an orbit around the
first or second charge.  If $\dist_{13}$ and $\dist_{23}$ change in
the same sense then the third charge will move toward or away from the axis
between the first and second charges.  For generic $k^I_j$, there will
only be a range of values for $\dist_{12}$ for which a solution is
possible.  The distances $\dist_{13}$ and $\dist_{23}$ can only
compensate for limited changes in $\dist_{12}$ without violating
triangle inequalities.

Perhaps the most instructive case to consider is when every $K$-charge
is equal: $k^I_i = k$, for all $i, I$,  $|q_j| =1$ for all $j$, and $q_0 =1$.
Decompose the Gibbons-Hawking points, $q_j$ into two sets:
\eqn\Qsets{ \cS_\pm ~\equiv~ \{j:  \ q_j = \pm 1 \} \,,}
Define the electrostatic potentials:
\eqn\Epots{\cV_\pm (\vec y) ~\equiv~ \sum_{j \in \cS_\pm} \, {8 \, k^2 \over 3}\, 
{1 \over |\vec y - \vec y^{(j)} |} \,.}
Then the equations \linsys\ reduce to:
\eqn\equipots{\cV_+  (\vec y^{(i)})  ~=~    (N+1) \,, \quad \forall \ i \in \cS_-\,;
\qquad  \cV_-  (\vec y^{(i)})  ~=~     (N-1) \,, \quad \forall \ i \in \cS_+ \,.}
These also have the redundancy arising from \redund:
\eqn\newred{\sum_{i \in \cS_-} \,\cV_+  (\vec y^{(i)}) ~=~ 
\sum_{i \in \cS_+} \,\cV_- (\vec y^{(i)}) \,.}
Thus we see that the positive Gibbons-Hawking points must be on a specific 
equipotential of $\cV_-$, and do not care where the other positive
charges are.  Similarly,  the negative Gibbons-Hawking points must be on a specific 
equipotential of $\cV_+$, and do not care where the other negative
charges are.

\goodbreak\midinsert
\vskip .2cm
\centerline{ {\epsfxsize 1.6in\epsfbox{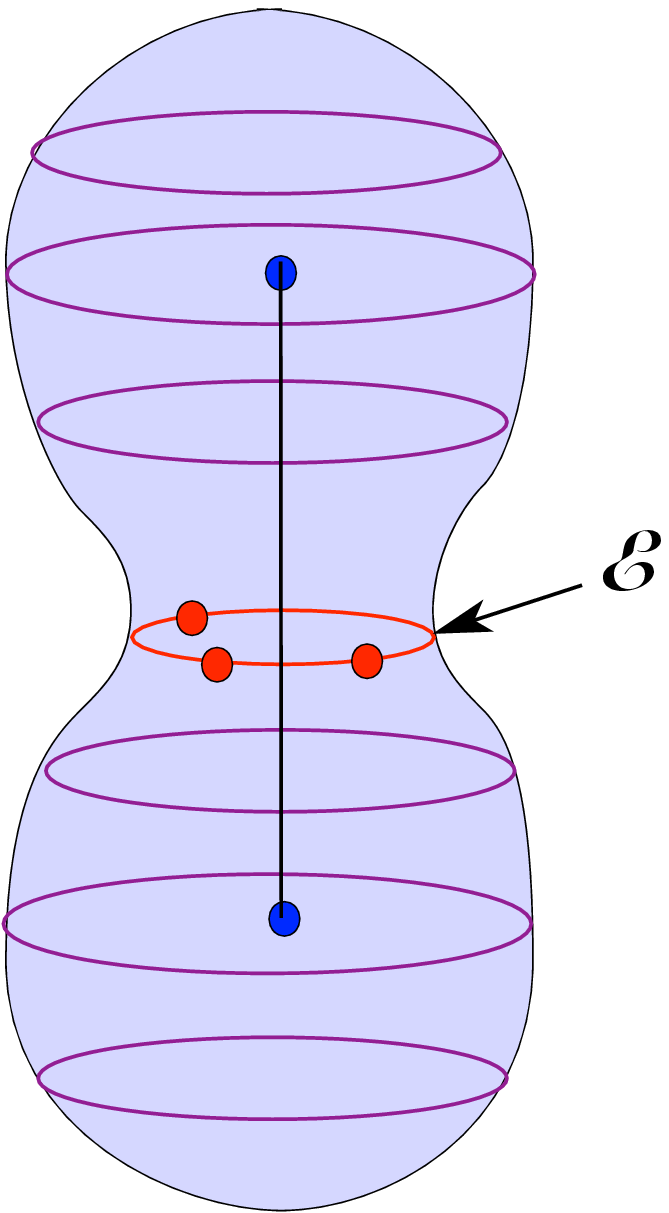}}}
\vskip 0.2cm
\leftskip 2pc
\rightskip 2pc\noindent{\ninepoint\sl \baselineskip=8pt
{\bf Fig.~2}:  This shows the equi-potential, $\cV_- ~=~ 4$. 
The three positive charges can be located anywhere on the equator, $\cE$. }
\endinsert

Suppose we have $N=3$ with labeling  \threeq, then \equipots\ tells
us that
\eqn\threesol{\dist_{13} ~=~  \dist_{23} ~=~  {4 \, k^2 \over 3} \,,}
and that $\dist_{12}$ is a free parameter.  However, the triangle inequality
limits $\dist_{12}$ to $0 \le \dist_{12} \le {8 \, k^2 \over 3}$.  

For $N=5$ things are a little more complicated.  Let $a$ be the separation
of the two negative charges.  Having fixed $a$ there should be a five-parameter
family of solutions to \equipots:  One must locate the positive charges
on the equi-potentials of $\cV_-$, which leads  to six parameters. 
There are then apparently two constraints coming from the first equation
\equipots\ but one of them is redundant via \newred.

There are some obvious solutions that can be obtained using
symmetry.   The potential, $\cV_-$, has a symmetry axis,
$\cA$, through the two Gibbons-Hawking points 
and the equi-potentials come either as a single surface, $\cB$, of revolution
about $\cA$, or it consists of two disconnected deformed spheres about each point.
If the two negative charges are close enough together then the equi-potentials
$\cV_- = (N-1) = 4$ is a single surface, $\cB$, which has a well defined
equator, $\cE$, mid-way between the two negative charges. (See Fig. 2.)
One obvious three-parameter solution to \equipots\ is to put all three positive
charges anywhere on $\cE$. The remaining two parameters come from
moving the charges off $\cE$:  One can move two of them in any way one 
wishes, and the third one's position is fixed by the first equation in
\equipots.  As for $N=3$, there will be limits on the range of motion
of the three points. 

Finally, we note that for this example, \VZIns\ collapses to yield the 
condition:
\eqn\VZsimp{ V \, Z_I ~=~\coeff{3}{8\, k^2}\,  \big(\cV_+ ~-~ \cV_-
~+~ \coeff{3}{2}\, \cV_+ \, \cV_- \big) ~\ge~ 0\,.}
For $N=3$ this inequality is simply:
\eqn\VZtriv{\eqalign{ {1 \over r_1}~+~  & {1 \over r_2}~-~  {1 \over r_3}~+~  
{4\, k^2 \over r_3} \, \Big({1 \over r_1}~+~  {1 \over r_2} \Big)  \cr 
& ~=~ {1 \over r_1}~+~  {(r_3 - r_2 + r_{23}) \over r_2 \, r_3} ~+~  
{ 1  \over r_3} \, \Big({4\, k^2 \over r_1}~+~  {(4\, k^2 - r_{23}) \over r_2} \Big) 
~\ge~0\,.}}
This is trivially satisfied because of the triangle inequality 
$r_3+ r_{23} \ge r_2$ and because  $r_{23}  = {4\, k^2 \over 3}$.

\newsec{Bubbling supertubes}

\subsec{The resolved solution}

In this section we investigate the form of the solutions that resolve
the singularity of the three-charge black ring.  We are also going to
require the resolved solution to have the same $U(1) \times U(1)$
symmetry as the black ring.  The geometric resolution we describe here
is depicted in Figure 1.

As discussed in Section 3, the metric on the base is given by
\eqn\bubble{V = {1 \over r} - {Q \over r_a} + {Q \over r_b}}
where $r_a$ and $r_b$ denote the distance to the points $a$ and
$b$.  If $a,\,b$ and the origin are colinear then the solution has a
$U(1)$ invariance from the $\IR^3$ of the GH metric and 
another $U(1)$ invariance from the GH fiber.

We want the non-singular solution outlined in section 4.  We can choose
the $K$-charges freely, but we only allow them to be sourced at the GH centers:
\eqn\bubk{K^I ~=~  {\kch_0^I \over r} + {\kch_a^I \over r_a} 
+ {\kch_b^I \over r_b} \,.}
Note that $\kch_0^I$ now denotes the charge at $r=0$, and not
the additive constant in \KLMexp.   As before,  we want this solution to be pure 
geometry with fluxes, with charges coming from fluxes, and
we require the harmonic functions $Z_i$ to have no
divergences at the points $0,a,b$. Together with asymptotic flatness, 
this completely determines the functions $L_I$:
\eqn\bubl{L_I ~ =~  1~+~ \half\, C_{IJK} \,\bigg(-{\kch_0^J \kch_0^K \over r} 
+ {\kch_a^J \kch_a^K \over Q r_a} 
- {\kch_b^J \kch_b^K \over Q r_b}\bigg)\,,}
where, as before, $C_{IJK} =C^{IJK} \equiv  |\epsilon_{IJK}|$.
Similarly, requiring $\mu$ to be regular leads to:
\eqn\bubm{M = \coeff{1}{12}\,   C_{IJK} \, \bigg({  \kch_0^I \kch_0^J \kch_0^K \over r} 
+ { \kch_a^I \kch_a^J \kch_a^K  \over  Q^2 \, r_a} 
+ {\kch_b^I \kch_b^J \kch_b^K \over  Q^2 \, r_b}\bigg) ~-~
\half \, \sum_{I,j=0,a,b} \kch_{j}^I  \,. }
To remove Dirac-Misner strings and CTC's leads to 
 three more relations of the form \linsys\ and only two of these relations 
 are independent.   Since we are taking the centers to be colinear, these
 equations determine $a$ and $b$ as a function of the  $\kch_j^I$. 
Thus the complete solution is fixed by the choice of the $\kch_j^I$, 
 which also fix the fluxes through the two $S^2$'s of the base.
 We will give these equations below but, as we noted in section 4,
 these equations are equivalent to requiring that $\vec \omega$ 
 can be written solely in terms of the $\vec \omega_{ij}$ of \omegaijdefn.
Here we have\foot{The solution in a more complicated $U(1)^n$ 
supergravity theory is obtained by 
simply replacing the products of the form
$\prod_{I=1}^3 \,( {k^I_a + k^I_b}) $ with ${1 \over 6} 
C_{IJK}({k^I_a + k^I_b})({k^J_a + k^J_b})( {k^K_a + k^K_b})$.}:
\eqn\omegabub{\vec \omega ~=~ 
\coeff{1}{2\, Q^2} \, \prod_{I=1}^3 \, ( \kch_a^I +Q\, \kch_0^I ) \, 
\vec \omega_{0a}  
~-~ \coeff{1}{2\, Q} \,  \prod_{I=1}^3 \, ( \kch_a^I + \kch_b^I )  \, \vec \omega_{ab}
~+~ \coeff{1}{2\, Q^2}  \,\prod_{I=1}^3 \,  
( \kch_b^I - Q \kch_0^I ) \, \vec \omega_{0b}  \,. }
%

\subsec{Orthogonal cycles and diagonalization}

As we have discussed in section 3, our solutions have a gauge
invariance \gauge, and therefore only two combinations of the three
parameters in $K_I$ appear in the solution. This gauge invariance
could be used to eliminate the $\kch_0^I$ for example; however, in
order to simplify the solution it is better to introduce other
variables that make the relation between the bubbling solutions and
the supertubes more direct.

To avoid unnecessary normalization conventions between the charges 
in the supergravity solution and the number of branes, we work in a 
convention where they are equal; this happens when the three $T^2$'s have equal
size and $G_5 = {\pi \over 4}$ \refs{\ElvangDS,\BenaNI}. The three M2-brane charges
are then
\eqn\charge{N_K ~=~   2\, C_{KIJ} \,  \Big(\Big(1 + {1 \over Q}\Big)\, 
\kch_a^I \,\kch_a^J +   2\, \kch_a^I \, \kch_b^J   +\Big(1 - {1 \over Q}\Big)\,
\kch_b^I \, \kch_b^J  +2\, \kch_a^I \, \kch_0^J  +  2\, \kch_b^I\, \kch_0^J \Big)\,.}
If we now introduce new, physical variables $n_I$ and $f_I$: 
\eqn\dia{ n_I ~\equiv~ 2 \,( \kch_a^I ~+~ \kch_b^I) \,, \qquad
 f_I  ~\equiv~  \coeff{1}{Q}\, \big[\kch_a^I \, (1+Q)+ \kch_b^I \,(Q-1) + 
 2\, Q\, \kch_0^I \big]\,,}
the charges become
\eqn\chargedia{N_K ~=~ C^{KIJ} \, n_I \, f_J\,.}
As expected from \gauge, all the 
components of the solution only depend on the $f$'s and $n$'s.
The $n_I$ and $f_I$ are simple combinations of fluxes through
the non-trivial cycles.  Indeed,  $n_I$ is proportional to the flux through
the cycle between $a$ and $b$ that resolves the supertube.  
The interpretation of $f_I$ is a bit more obscure; however,  when  $Q=1$, 
then  $ n = 2(\kch_a + \kch_b)$ and $ f = 2(\kch_a + \kch_0)$, and so the 
cycles that give $n$ and $f$ run between the point with a minus GH charge 
and the two points with positive GH charge.

\subsec{Solving the bubble equations}

Since we are restricting the GH centers to lie on an axis there
are, {\it a priori},  several  distinct cases determined by the
signs of $a,b$ and $b-a$.   However,  for the bubbling supertube
we will only need consider the  regime  with  $b>a>0$.    
The bubble equations are:
\eqn\bubbleone{ \eqalign{{1\over a \, Q^2}\,  \prod_{I=1}^3 \,  \big( \half\, (Q -1 ) \, 
{n_I} -  Q\, {f_I} \big )   ~-~ {1\over b \, Q^2}\,  \prod_{I=1}^3 \,  \big(\half\, (Q +1 ) \, 
{n_I} &  -    Q\, {f_I} \big )\cr & ~+~ 4\, \sum_{I=1}^3 \, {n_I} ~=~ 0 \,, }}
\eqn\bubbletwo{ \eqalign{ {1 \over (b-a)\,Q} \,\prod_{I=1}^3 \, {n_I}  ~-~ {1\over b \, Q^2}\,  
 \prod_{I=1}^3  \, \big(\half \, (Q+1)\, {n_I} & -  Q\, {f_I} \big)\cr    ~-~&  
 4\, \sum_{I=1}^3 \, \big( \half\, (Q-1) {n_I}  +   Q {f_I}  \big)   ~=~ 0\,.}}
These equations are also trivially generalized to the case of a more
complicated $U(1)^n$ supergravity theory.

The general solutions of these  equations are  quite involved,  however they
have some  rather interesting  features. 
The simplest case to analyze is to take $|Q|=1$ and $n_I = f_I = 4 k$  for all $I$
as we did in the previous section.   For $Q=1$ we learn from \threesol,
or \bubbleone, \bubbletwo\  that
\eqn\colinthreea{a ~=~  {4\, k^2 \over 3} \,, \qquad b  ~=~ {8\, k^2 \over 3}\,,}
thus the negative
GH charge is exactly in the middle of the two positive charges.  For 
$Q= -1$ we have:
\eqn\colinthreeb{a ~=~  0\,, \qquad b  ~=~ {4\, k^2 \over 3} \,,}
and so the two positive GH charges coincide at the origin.

Of more interest physically is the limit in which the distance between
$a$ and $b$ becomes very small compared with the distance from the
origin to the points $a$ and $b$. As we will see 
in the next subsection, the latter distance asymptotes to
the radius of the naive supertube, and the solution
resembles the naive supertube solution.  The distance between $a$ and
$b$ is given by the balance between the attraction of the $Q$ and $-Q$
charges, and the tendency of the fluxes wrapped on the two-cycle between $a$ and $b$ 
to expand this cycle and make this distance larger. 
Therefore, one expects to obtain a solution that matches the
naive supertube solution {\it both} in the small flux limit (small
$n$) and in the large $Q$ limit. 
 
When $a$ and $b$ are very close to each other it is relatively easy to
find an approximate solution to \bubbleone\ and \bubbletwo.   Equation
\bubbletwo\ determines the separation between $a$ and $b$, and equation
\bubbleone\ gives the separation between these two points and the
origin. The leading part of \bubbleone\ (obtained by setting $a=b$)
then gives:
\eqn\solbubble{
a \approx b \approx  {  2\, C^{IJK}\,( {f_I}\,{f_J}\,{n_K} - {n_I}\,{n_J}\,{f_K} )
+ \big(3+{1 \over Q^2}\big) \coeff {1}{6} C^{IJK}\, {n_I}\,{n_J}\,{n_K}  \over 16 \, \sum_{I=1}^3 {n_I}  } }
As we will see in the next subsection, this matches the radius of the naive
supertube solution both in the small $n$ and in the large $Q$ limits.
One can also estimate the size of the bubble, $a-b$; we will give
the complete formula for this size when we discuss bubbling supertubes in 
Taub-NUT in subsection 5.5.

We have also numerically checked that $\cQ > 0$ and  $r^2 \sin^2 \theta \, d \phi^2 -{\omega^2
\over \cQ} \ge 0$  in an example in which 
$a$ and $b$ are large compared to $a-b$. Thus, the bubbling of the supertube 
does not generate CTC's.

The angular momenta of the bubbling supertube are easy to find.  The
complicated form of the solutions of the bubble equations 
might have worried one that the
angular momenta, which depend explicitly on $a$ and $b$ would be rather
horrible. However, a pleasant surprise awaits.

To read off the angular momenta we first do a change of coordinates to write
the asymptotic form of the GH base as $\IR^4$. If one chooses asymptotically
$\vec{A} \cdot d\vec{y} = (1+ \cos \theta ) d \phi$ in \GHmet, then the 
coordinate change:
\eqn\eb{  \phi  ~=~ \phit_2 - \phit_1 \,, \qquad \psi ~=~ 2\, \phit_1\,,
\qquad r ~=~ \coeff{1}{4} \, \rho^2 \,.}
makes the  $\IR^4$ form of the asymptotic solution explicit. The two 
angular momenta are then determined from the 
asymptotic behavior of $\mu$ and $\omega$ via: 
\eqn\joneinf{\mu (1-\cos \theta) - \omega 
\approx {J_1 \sin^2{\theta\over 2}\over 4 r}}
\eqn\joneinf{\mu (1+\cos \theta) + \omega 
\approx {J_2 \cos^2{\theta\over 2}\over 4 r}}

One combination of the 
angular momenta is independent of $a$ and $b$, and is given by
\eqn\jplus{ \, (J_1+J_2) ~=~ 
 - { (Q^2  - 1)   \over 4\, Q^2}\, 
\coeff {1}{6} C^{IJK}\, {n_I}\,{n_J}\,{n_K}
~+~ \half \,
C^{IJK}\, \big( {n_I}\,{n_J}\,{f_K} ~+~  {f_I}\,{f_J} \,{n_K} \big)\,. }

Despite the very complicated form of the solution of \bubbleone\ and 
\bubbletwo, the other combination of the angular momenta 
is also very simple. The final results are:
\eqn\jone{   J_1 ~=~  - {  (Q   - 1) \over 2\, Q} \, 
\coeff {1}{6} C^{IJK}\, {n_I}\,{n_J}\,{n_K}
~+~ \half \, C^{IJK}\, {n_I}\,{n_J}\,{f_K} \,,}
\eqn\jtwo{ J_2 ~=~  {{ ( Q -1 ) }^2 \over 4 \, Q^2} \,
\coeff {1}{6} C^{IJK}\, {n_I}\,{n_J}\,{n_K}
~+~ \half \,C^{IJK}\, {f_I}\,{f_J} \,{n_K} \,.}
As we will see in the next subsection, these will again 
match the naive supertube angular momenta, 
both in the large $Q$ limit and in the small $n_I$ limit.

\subsec{Matching the naive supertube solution}

We first write the naive solution describing the zero-entropy black ring 
by rewriting the $\IR^4$ base as a GH metric with a single
GH center of charge one.  This single center can be taken to be at $r=0$,
and then the supertube generically has sources for $K,L$ and $M$ at
$r=0$ and at $r_a =0$.  The radius, $R$, of the supertube is then 
the distance, $r_{0a}$, between the two source points.  We can use 
the gauge invariance, \gauge, to set the $K$-charge at $r=0$ to
zero, and then the supertube solution is given by \BenaNI:
\eqn\tubesol{\eqalign{  V ~=~ {1 \over r}  \,, \qquad K^I  
~=~ {n_I \over 2 \, r_a} \,, 
\qquad L_I  ~=~ 1+{\Nb_I \over 4 \, r_a}\,, \qquad
M  ~=~ - {J_T \over 16}\, \bigg({1 \over R}- {1 \over r_a}\bigg) \,. }}
In this expression,  $n_I $ and $\Nb_I$ are the (integer) numbers of
M5-branes and M2-branes that make up the supertube, and $J_T$
is the angular momentum of the tube alone. The ``tube'' angular momentum, 
$J_T$ is related to the radius by 
\eqn\JTRreln{J_T ~=~  4 \, R_T\, (n_1 + n_2 + n_3) \,,}
In the case of the zero-entropy black ring, this angular momentum
is completely determined in terms of the  $n_I $ and $\Nb_I$:
\eqn\entropy{J_T = {2\, n_1\, n_2 \Nb_1 \Nb_2  +2\, n_1\, n_3 \,\Nb_1 \Nb_3  
+2\, n_2\, n_3\, \Nb_2 \Nb_3 - n_1^2\, \Nb_1^2 - n_2^2\, \Nb_2^2 
- n_3^2\, \Nb_3^2 \over 4\, n_1 \, n_2\, n_3 }, }
where $\Nb_1 = N_1 - n_2 n_3$, and similarly for  $\Nb_2$ and $\Nb_3$. 
The angular momenta of the naive supertube solution are
\eqn\jflux{\eqalign{
J_1 &=  {n_1 \Nb_1 + n_2  \Nb_2 + n_3 \Nb_3 \over 2 } + n_1 n_2 n_3 \cr
J_2 &= J_T+ {n_1 \Nb_1 + n_2  \Nb_2 + n_3 \Nb_3 \over 2 } + n_1 n_2 n_3 
}}

As one can see both from the form of the $K^I$ when $a$ and $b$ are very close, and
from the integral of the fluxes on the cycle that runs between $a$ and $b$, the 
$n_I$ of the bubbling solutions are identical to the M5 dipole charges $n_i$.
If one then interprets \chargedia\ as a change of variables between the 
$N_I$ and the $f_I$, one can express the supertube angular momenta and radius 
in terms of the $n_I$ and $f_I$:

\eqn\entropytwo{J_T = \half\, C^{IJK}\,( {f_I}\,{f_J}\,{n_K} 
- {n_I}\,{n_J}\,{f_K} )
+ \coeff {1}{8} C^{IJK}\, {n_I}\,{n_J}\,{n_K}
\,,}
\eqn\jfluxtwo{\eqalign{J_1 &= \half \, C^{IJK}\,( {n_I}\,{n_J}\,{f_K} )
 ~-~ \coeff {1}{12} C^{IJK}\, {n_I}\,{n_J}\,{n_K} \,,\cr
J_2 &= \half \, C^{IJK}\,( {f_I}\,{f_J}\,{n_K} )
 ~+~ \coeff {1}{24} C^{IJK}\, {n_I}\,{n_J}\,{n_K}
}}
which gives
\eqn\rnaive{R_T = {  2\, C^{IJK}\,( {f_I}\,{f_J}\,{n_K} - {n_I}\,{n_J}\,{f_K} )
+  \coeff {1}{2} C^{IJK}\, {n_I}\,{n_J}\,{n_K}
\over 16 \,  \sum_{I=1}^3 {n_I}   }  .} 

As we have advertised in the previous subsection, these match exactly the
 bubbling supertube radius and angular momenta both in
 the limit when the dipole charges
$n_I$ are small, and in the limit when $Q$ is large.

We should also note that the $n\, n \, f$ and $n \, f \, f$ combinations that appear 
in the angular momentum formulae \jone\ and
\jtwo\ are not apparent at all from the form of the supertube angular
momenta, and only become apparent after expressing the $N_I$ using
\chargedia. When $Q=1$ the angular momenta are interchanged under the
exchange of $n_I$ and $f_I$. Hence, a solution with $Q=1$ has two interpretations:
for small $n_I$ it is a supertube of dipole charges $n_I$ in
the $\phit_1$ plane, and for small $f_I$ it is a supertube of dipole
charges $f_I$ in the $\phit_2$ plane. This feature -- the existence of  
one supergravity solution with two different brane interpretations -- is present
in all the other systems that contain branes wrapped on topologically trivial cycles 
\refs{\PolchinskiUF,\BenaQV,\LinNB} and might be the key to finding the 
microscopic description of our bubbling supertube geometries.

When the size of one bubble is much smaller than the size of the other,
the harmonic functions that give the bubbling solution \bubk, \bubl\ and \bubm\
become approximately equal to the supertube harmonic functions \tubesol. 

If we work in the gauge with $k_0^I=0$, one can easily see that 
\bubk\ combined with \dia\ reproduces the correct identification of
the dipole charges with the flux integrals. 
From \bubl\ one finds:
\eqn\Nbident{\Nb_I  ~=~   \coeff{2}{Q}\, C_{IJK}\, \big(k^J_a \,k^K_a ~-~ 
k^J_b \,k^K_b \big)  \,.}
This again agrees with \chargedia\ and \dia\ in the gauge $k_0^I=0$ 
after using the fact that the 
total M2-brane charge of the solution is
\eqn\NbNrel{\eqalign{N_I  &~=~  \Nb_I ~+~  \coeff{1}{2}\, C^{IJK}\,  n_J\, n_K \cr
& ~=~  \coeff{2}{Q}\, C_{IJK}\, \Big( (k^J_a \,k^K_a ~-~ 
k^J_b \,k^K_b \big)~+~  Q\,  ( \kch_a^J+ \kch_b^J)\, 
( \kch_a^K+ \kch_b^K) \Big) \cr
&  ~=~  \coeff{2}{Q}\, C_{IJK}\,   ( \kch_a^J+ \kch_b^J)\,  
( (Q+1) \, \kch_a^K+ (Q-1) \,  \kch_b^K)\,  \,.}}

Finally, the harmonic function $M$ in \bubm\  reproduces that of
the naive supertube after identifying
\eqn\JTres{\eqalign{J_T ~=~ & \coeff{4}{3 \, Q^2} \, C_{IJK}\, (k_a^I \, k_a^J
\,k_a^K ~+~  k_b^I \, k_b^J \,k_b^K )\cr  ~=~ & \half\,  C^{IJK}\,
( {f_I}\,{f_J}\,{n_K} - {n_I}\,{n_J}\,{f_K} ) ~+~ \coeff{1}{24}\, 
\Big(3  - {4\over  Q} + {1 \over Q^2}\Big) 
 C^{IJK}\, {n_I}\,{n_J}\,{n_K}
}}
and remembering that the solutions agree in the large $Q$ or small $n$ limit.

Hence, the bubbling solution is identical to the naive solutions at
distances much larger then the size of the bubble. Moreover, in the limit when $Q 
\rightarrow \infty$ or in the limit when $n_I \rightarrow 0$ with
fixed $N_I$, the bubble that is nucleated to resolve the singularity
of the three-charge supertube becomes very small, and the resolved
solution becomes virtually indistinguishable from the naive
solution. This confirms the intuition coming from the discussion of
geometric transitions in section 2, and from the similar phenomenon
observed in \LinNB. 

The fact that for small dipole charges the singularity resolution is
local provides very strong support to the expectation that there exist
similar bubbling supertube solutions that correspond to supertubes of
arbitrary shape and arbitrary charge densities.

\subsec {Bubbling supertubes in Taub-NUT}
 
In order to compare our singularity resolution mechanism 
to that of the zero-entropy  four-dimensional black hole \BenaAY, 
we need to put the bubbling  supertube in Taub-NUT. This solution
resolves the singularity of the zero-entropy black ring in Taub-NUT  
\refs{\ElvangSA,\GaiottoXT,\BenaNI}.

The construction of bubbling supertubes in Taub-NUT is almost identical to
the construction in asymptotically Euclidean space. To simplify the algebra
we make use of the gauge freedom \gauge\ to set the three $\kch^I_0$
to zero. The asymptotically Taub-NUT base is obtained by modifying
the harmonic function $V$ to
\eqn\bubbletn{V = h ~+~ {1 \over r} ~-~ {Q \over r_a} ~+~ {Q \over r_b} }
where we do not fix $h$ to one  in order to make the interpolation between
the asymptotically $\IR^4$ and the asymptotically   Taub-NUT solutions easier.
The functions $K^I$ and $L_I$ are the same as before (see   
\bubk\ and \bubl). The coefficients $m_j$ in $M$ are  given by 
the requirement that $\mu$ be regular at the three centers \mchchoice, 
and they are also not changed. However, because of the changed asymptotics 
there is no longer a requirement that $\mu$ vanish at infinity, and so 
$m_0$ is a free parameter. Hence, 
\eqn\bubmtn{M ~=~ m_0 ~+~  \coeff{1}{12}\, C_{IJK} \, \bigg( {  \kch_0^I 
\kch_0^J  \kch_0^K  \over  r}  + { \kch_a^I \kch_a^J \kch_a^K  \over Q^2\, r_a} 
+ {\kch_b^I \kch_b^J \kch_b^K \over  Q^2\, r_b}\bigg) \,. }

The requirement that $\mu$ vanishes at the three centers gives three
equations, which are now independent:
\eqn\bubonetn{ \eqalign{{1\over a \, Q^2}\,  \prod_{I=1}^3 \,  \big(\half\, (Q -1 ) 
\, {n_I} -   Q\, {f_I} \big )   ~-~ {1\over b \, Q^2}\,  \prod_{I=1}^3 \,  \big( \half\,
(Q +1) \, {n_I} &  -  Q\, {f_I} \big )\cr & - 16\, m_0 ~=~ 0 \,,
}}
\eqn\bubtwotn{ \eqalign{ {1 \over (b-a)\,Q} \,\prod_{I=1}^3 \, {n_I}   ~-~ 
\bigg(h +{1 \over b}\bigg) & {1\over \, Q^2}\,  \prod_{I=1}^3  \, 
\big(\half \, (Q+1)\, {n_I}  -  Q\, {f_I} \big)\cr  ~-~  & 16\, m_0  ~+~  
\sum_{I=1}^3 \, \big(\half\, (Q+1) {n_I}  -   Q {f_I}  \big)   ~=~ 0\,.}}
\eqn\bubthreetn{ \eqalign{ {1 \over (b-a)\,Q} \,\prod_{I=1}^3 \, {n_I}  ~-~ 
\bigg(h +{1 \over a}\bigg) & {1\over \, Q^2}\,   \prod_{I=1}^3  \, \big(\half\, 
(Q-1)\, {n_I}  -  Q\, {f_I} \big)\cr  ~+~& 16\,m_0\, Q ~+~  \sum_{I=1}^3 \, 
\big(\half\, (Q-1) {n_I}  -   Q {f_I}  \big)   ~=~ 0\,.}}

The compatibility of these equations determines $m_0$ to be
\eqn\mtn{m_0 ~=~ \coeff{1}{32} \,h\, C^{IJK}\,( {f_I}\,{f_J}\,{n_K} - {n_I}\,{n_J}
\,{f_K} )  ~+~ \coeff{1}{384}\, \Big(3+{1 \over Q^2}\Big) \,h\, 
 C^{IJK}\, {n_I}\,{n_J}\,{n_K} 
~-~\coeff{1}{4}\,\sum_{I=1}^3 {n_I} \,.}
When $h =0$ the value of $m_0$ becomes $-  {1 \over 4} \sum_{I=1}^3 \, {n_I}$, 
and one recovers the solution of the previous subsection in the gauge  $k^I_0 = 0$.

One can solve these equations in the large $Q$ or in the small $n_I$
limit, in which the separation between $a$ and $b$ is much smaller than
their distance from the origin. Setting  $a = b$ in  equation \bubonetn,
one obtains:
\eqn\radiustn{ h + {1 \over a} \approx  h_0 \equiv {16\, \sum_{I=1}^3 {n_I} 
\over 2\, C^{IJK}\,( {f_I}\,{f_J}\,{n_K} - {n_I}\,{n_J}\,{f_K} )
+ \Big(3+{1 \over Q^2}\Big) 
\coeff {1}{6} C^{IJK}\, {n_I}\,{n_J}\,{n_K}
} \,. }
In the large $Q$ limit, this equation reproduces the correct radius of
the zero-entropy black ring in Taub-NUT \BenaNI.

To reach the four-dimensional black hole limit one needs to move the
bubbling supertube away from the Taub-NUT center, to very large $a$
and $b$, keeping the fluxes $n_I$ and $f_I$ fixed.

To do this, one adjusts $h$ until it reaches $h_0$. 
In this limit one has $m_0 = 0$, and \bubonetn\ is trivially satisfied.
The distance between $a$ and $b$ is obtained from either of the remaining two
equations:
\eqn\diff{b-a \approx  { Q \,\prod_{I=1}^3 \, {n_I}  \over  {h_0}\,  
 \prod_{I=1}^3  \, \big(\half\, (Q-1)\, {n_I}  -  Q\, {f_I} \big)  
~-~ Q^2 \sum_{I=1}^3 \, \big(\half\, (Q-1) {n_I}  -   Q {f_I}  \big)}  \,, }
where $h_0$ is defined in \radiustn. 

Thus, by putting the supertube in Taub-NUT we have related the
singularity resolution mechanism of the zero-entropy black ring (the
nucleation of two oppositely-charged GH centers) to the singularity
resolution of the zero-entropy four-dimensional black hole (the
splitting of the branes that form the black hole into two stacks at a
finite radius from each other \BenaAY). Similar solutions
have also been analyzed from the point of view of four-dimensional
supergravity in \refs{\DenefNB\BatesVX{--} \DenefRU}.

We see therefore that the interpolation between Taub-NUT and $\IR^4$ is not 
only a good tool to obtain the microscopic description of black rings
and black holes \refs{\GaiottoGF,\BenaNI}, but is also useful in 
understanding their singularity resolution.

\newsec{Conclusions}

We have constructed smooth geometries that resolve the zero-entropy
singularity of BPS black rings. The $U(1) \times U(1)$ invariant
geometries must have a Gibbons-Hawking base space with several centers
of positive and negative charge. Despite the fact that the signature
of the base changes, the full geometries are regular.

These geometries stem naturally from implementing the mechanism of
geometric transitions to the supertube. The physics is closely related
to that of other systems containing branes wrapped on topologically
trivial cycles \LinNB.  The fact that our solutions reproduce in one
limit the geometries found by Mathur and collaborators
\refs{\GiustoID,\GiustoIP,\GiustoKJ}, and in another limit they reduce
to the ground states of the four-dimensional black hole found in
\BenaAY, is a non-trivial confirmation that these geometries are
indeed the correct black ring/supertube ground states.

The BPS black rings have two microscopic interpretations: one in terms of the 
D1-D5-P CFT \BenaTK, and another in terms of a four-dimensional black hole CFT 
\refs{\BenaTK,\BenaNI,\CyrierHJ}.     Hence, our solutions should similarly 
be thought of as {\it both} microstates of the D1-D5-P system, and as
microstates of the four-dimensional black hole CFT that describes the
black ring \refs{\BenaTK,\BenaNI}.  To establish the four-dimensional
black hole microstates that are dual to our solutions it is best
perhaps to put the bubbling supertubes in Taub-NUT, and go to the
limit when they become four-dimensional configurations.  However, to
establish their interpretation in the D1-D5-P CFT is somewhat
non-trivial.  One possibility is to start from the D1-D5-P microscopic
description of the BPS black ring \BenaTK, and to explore the
zero-entropy limit.  However, this might not be so straightforward,
since the naive supertube and resolved geometries only agree in the
very large $Q$ limit.  Another option is to use the fact that when the
two positively charged centers coincide these solutions reproduce
those of \refs{\GiustoID,\GiustoIP,\GiustoKJ}, which do have a D1-D5-P
microscopic interpretation.

Our analysis also indicates that the most generic bound state with a GH
base is determined by a gas of positive and negative centers, with
fluxes threading the many non-trivial two-cycles of the base, and no
localized brane charges. This proposal has similar features to the
foam described in \IqbalDS, but in \IqbalDS\ the foam was restricted
to the compactification space, whereas here the foam naturally lives
in the macroscopic space-time and defines the interior structure of a
black hole. Another interesting exploration of a similar type of spacetime foam
from the point of view of a dual boundary theory has appeared in \vijay.

Our results suggest quite a number of very interesting consequences
and suggestions for future work.  First, we have only considered
geometries with a Gibbons-Hawking base, because such geometries are
easy to find \refs{\GauntlettQY,\GauntlettNW}, and appear in the $U(1)
\times U(1)$ invariant background.  However, the most general smooth
solution will have a base given by an asymptotically $\IR^4$
hyper-K\"ahler manifold whose signature can change from $(+,+,+,+)$ to
$(-,-,-,-)$. We expect these solutions to give regular geometries\foot{It is also 
worth stressing that it is the possibility of signature change that 
enables us to avoid the extremely restrictive conditions, familiar to relativists,
on the existence of four-dimensional, asymptotically Euclidean
metrics.  By allowing the signature of the base to change we have
found a large number, and conjectured an even larger number of base
spaces that should give smooth three-charge geometries.},
provided that there are non-zero dipole fluxes.

The classification of the generalized hyper-K\"ahler manifolds that
we use here is far from known.  However, one may not need the metrics 
to do interesting
physics.  We are proposing that the black hole microstates are
described by a foam of non-trivial $S^2$'s in a four-dimensional base.
One might be able to do some statistical analysis of such a foam,
perhaps using toric geometry, to see if one can describe the
macroscopic, bulk ``state functions'' of the black hole. It is also
interesting to investigate the transitions between different
geometries, nucleation of GH points, instantons for such transitions,
and probabilities of transition.  Presumably nucleation is easy for
small fluxes and small GH charges, but there should be some kind of
correspondence limit in which large, classical bubbling supertubes, which
involve only two GH points with very large $Q$'s, should be relatively
stable and not decay into a foam.

It is interesting to note that ideas of space-time foam have made
regular appearances in the discussion of quantum gravity, see, for
example \HawkingZW.  In a very general sense, what we are proposing
here is in a similar spirit to the ideas of \HawkingZW: Space-time on small
scales becomes a topological foam.  Here, however, we have managed to
find it as a limit of supersymmetric D-brane physics, and with this
comes a great deal more computational control of the problem.  It is
also important to note that the same physical ideas that led to the
idea that space-time becomes foamy near the Planck scale also come
into play here.  At the Planck scale, even in empty space, there are
virtual black holes, and we are proposing that their microstates be
described by foams of two-spheres that will be hyper-K\"ahler only for
BPS black holes.  Consistency would therefore suggest that 
these virtual black holes should really be virtual fluctuations
in bubbling hyper-K\"ahler geometries. Therefore, even empty
space should be thought of in terms of some generalization of the
foamy  geometries considered here.    Obviously our
geometric description will break down at the Planck scale, but the
picture is still rather interesting, and it is certainly supported by the
fact that large bubbles are needed to resolve singularities in
macroscopic supertubes and black rings.

There are evidently many things to be tested and lots of interesting
things that might be done, but we believe that  we have made important 
progress by resolving the supertube singularity and thereby giving a 
semi-classical  description of black-hole microstates that
may also give new insight into the structure of space-time on very
small scales.

\bigskip
\noindent {\bf Acknowledgments:} \medskip \noindent
We would like to thank Eric Gimon, Jan Gutowski, Clifford Johnson, Per Kraus, 
Jim Liu,  Don Marolf,  Harvey Reall and Radu Roiban or interesting discussions. 
The work of IB is supported in part by the NSF grant  PHY-00-99590. The work of NW
is supported in part by the DOE grant DE-FG03-84ER-40168.

\listrefs
 \end